\documentclass{article}

\usepackage[english]{babel}
\usepackage[utf8x]{inputenc}
\usepackage{amsmath}
\usepackage{amssymb}
\usepackage{bbm}
\usepackage{amsthm}
\usepackage{graphicx}
\usepackage{listings}
\usepackage[lofdepth,lotdepth]{subfig}
\usepackage{fullpage}
\usepackage{multirow}
\usepackage{caption}
\usepackage{array}
\usepackage{wrapfig}
\usepackage{morefloats}
\usepackage{pdflscape}
\usepackage{natbib,bm}
\usepackage{longtable}
\usepackage{forest}
\usepackage{pdflscape}
\usepackage{booktabs}
\usepackage{rotating}
\usepackage{tcolorbox}
\usepackage{hyperref}
\usepackage{appendix}
\usepackage{titlesec}
\usepackage{authblk}

\begin{document}




\title{Collapsible Kernel Machine Regression for Exposomic Analyses}

\author[1]{Glen McGee }
\author[2]{Brent A. Coull}
\author[3]{Ander Wilson}

\affil[1]{Department of Statistics and Actuarial Science, University of Waterloo. ({\tt gmcgee@uwaterloo.ca})}
\affil[2]{Department of Biostatistics, Harvard T. H. Chan School of Public Health}
\affil[3]{Department of Statistics, Colorado State University}








\maketitle

\begin{abstract}
    
An important goal of environmental epidemiology is to quantify the complex health risks posed by a wide array of environmental exposures. In analyses focusing on a smaller number of exposures within a mixture, flexible models like Bayesian kernel machine regression (BKMR) are appealing because they allow for non-linear and non-additive associations among mixture components. However, this flexibility comes at the cost of low power and difficult interpretation, particularly in exposomic analyses when the number of exposures is large. We propose a flexible framework that allows for separate selection of additive and non-additive effects, unifying additive models and kernel machine regression. The proposed approach yields increased power and simpler interpretation when there is little evidence of interaction. Further, it allows users to specify separate priors for additive and non-additive effects, and allows for tests of non-additive interaction. 
We extend the approach to the class of multiple index models, in which the special case of kernel machine - distributed lag models are nested.   We apply the method to motivating data from a subcohort of the Human Early Life Exposome (HELIX) study containing 65 mixture components grouped into 13 distinct exposure classes.
\end{abstract}

\section{Introduction}

Characterizing the relationships between health outcomes and a wide array of environmental exposures---or exposomic analysis---is a major priority of the current and future strategic plan of the  National Institute of Environmental Health Sciences \citep{national20182023,national20252029}. This marks an ongoing shift from simpler, single-pollutant analyses and low dimensional multi-pollutant mixture analyses to large scale analyses that attempt to quantify the full burden of a person's environment (their exposome) \citep{wild2012exposome}. Among the many challenges in these environmental analyses are the facts that the functional form of exposure-reponse relationships is unknown and exposures may interact in complex ways \citep{carlin2013unraveling,taylor2016statistical,maitre2022state,Chung2024exposomedatascience}. A popular non-parametric approach proposed for mixtures analyses, Bayesian kernel machine regression (BKMR), addresses both issues, allowing non-linear exposure response relationships as well as high-order interactions in a parsimonious model \citep{bobb2015bayesian,bobb2018statistical}.
In practice, however, it is often too flexible. Coupling non-linearity with non-additivity in a kernel framework makes identifying either type of relationship challenging, particularly in small samples with small effect sizes common in environmental health research. Even when there is enough power to detect these effects, interpretation can be prohibitively difficult in exposomic analyses, where there is a large number of exposures ($p$), requiring investigation of each ($p$) individual exposure-response curve and each of $p\times (p-1)/2$ two-way interaction plots. 

In contrast to BKMR, some simpler approaches like exposome wide association studies (EWAS) have been proposed to scale to higher numbers of exposures \citep{agier2016systematic}, but these screening approaches are restrictive, not allowing for interactions or non-linearity. Instead recent work has extended BKMR  to reduce its flexiblity. Adapting a projection approach from the spatial literature, 
\cite{ferrari2020identifying} improved efficiency by projecting out linear effects and two-way (linear) interactions. This improved efficiency when most effects were linear but did not improve interpretability, and the nonidentifiability precluded hypothesis testing. \cite{antonelli2020estimating} circumvented the kernel framework and assigned hierarchical variable selection priors to a natural cubic spline basis expansion, but computation is challenging when there are a large number of exposures. Another approach suited to exposomic analyses is the Bayesian multiple index model (BMIM), which exploits group structure in the exposures. The BMIM reduces the dimension of the inputs of BKMR by constructing linear multi-exposure indices defined as weighted sums of exposures with weights estimated from the data \citep{mcgee2021bayesian,mcgee2022integrating}. 
However, when the number of exposures, or the number of indices in a multiple index model, is large, these approaches can be overly flexible and challenging to interpret.
 
We propose a collapsible kernel machine regression (CKMR) framework that decouples non-linear additive relationships from non-additive interactions. A novel adaptive projection approach decomposes the non-parametric surface into an additive function space modeled by basis expansions and its orthogonal complement that contains interactions. 
Paired with a hierarchical variable selection prior, the model can collapse to a generalized additive model (GAM) when there is no evidence of interactions. At the same time, the model framework allows for interactions among a subset of exposures or all exposures through the BKMR framework when this level of complexity is supported by the data.
This strategy has three advantages over non-parametric approaches like BKMR. First, it improves efficiency when there is little evidence of interactions. Second, it improves interpretability by obviating the need to investigate a large number of interaction plots. Third,  thanks to a novel adaptive projection approach, it allows one to test for non-additive interaction separately from overall tests of association between an exposure and the outcome. 

In this paper we analyze a motivating dataset from the Human Early Life Exposome (HELIX) project \citep{maitre2018human,vrijheid2014human}. In particular, interest lies in characterizing the complex relationships between youth BMI and 65 different environmental and chemical exposures, grouped into 13 distinct exposure classes. To that end we extend the  collapsible kernel framework to multiple index models, and apply the method to the HELIX data, treating each  exposure class as a separate index.

The rest of the paper proceeds as follows. In Section \ref{s:background}, we briefly review existing methods for mixtures analyses.  In Section \ref{s:ckmr} we present the novel collapsible kernel machine regression framework and collapsible index models. We then evaluate the performance of the collapsed methods in simulations in Section \ref{s:sims}, and apply it to motivating data on associations between youth BMI and 13 different classes of exposures from the Human Early Life Exposome (HELIX) study in Section \ref{s:analysis}. We conclude with a discussion in Section \ref{s:disc}.



\section{Background} \label{s:background}
\subsection{BKMR}  \label{ss:bkmr}
Let $y_i$ be a continuous outcome of interest,  $\mathbf{x}_i=(x_{i1},\cdots,x_{ip})^T$ be a vector of $p$  exposures and $\textbf{z}_{i}$ be a vector of covariates for the $i^{th}$ observation ($i=1,\cdots,N$). The BKMR model is
\begin{align*}
y_i &= h(\mathbf{x}_i)+\mathbf{z}_i^T\boldsymbol{\alpha} +\epsilon_i, ~~\epsilon_i\sim N(0,\sigma^2), 
\end{align*}
where $h(\cdot): \mathbb{R}^p \to \mathbb{R}$ is an unknown and potentially non-linear function of $\mathbf{x}_i$, and $\boldsymbol{\alpha} $ is a vector of regression coefficients. 
The exposure-response function $h(\cdot)$ is defined by a kernel function $K:\mathbb{R}^p \times \mathbb{R}^p \to \mathbb{R}$ \citep{cristianini2000introduction}.
By default we use a Gaussian kernel, $K(\mathbf{x}_i,\mathbf{x}_{i'})=\exp\left[-\sum_{j=1}^p \rho_j (x_{ij}-x_{i'j})^2\right],$ where  $\rho_j \geq 0$ is an unknown component weight; this corresponds to a radial basis function representation of  $h(\cdot)$.

The model can be conveniently represented as a linear mixed effects model \citep{liu2007semiparametric}:
\begin{align}
y_i|h_i &\sim N(h_i+\mathbf{z}_i^T\boldsymbol{\alpha},\sigma^2),  \label{eqn:lme}\\
(h_1,\cdots,h_N)^T&\sim N(\mathbf{0},\nu^2\sigma^2  \mathbf{K}),\nonumber 
\end{align} 
where $\mathbf{K}$ is the kernel matrix with elements $\mathbf{K}_{ij}=K(\mathbf{x}_i,\mathbf{x}_j)$, and $\nu^2>0$ is a penalty term that controls smoothness. We then base estimation and inference on the marginal likelihood for $\mathbf{y}$.  The model is completed by specifying priors for $\{\boldsymbol{\rho},\boldsymbol{\gamma},\sigma^2,\nu^2\}$, and we adopt  default priors from  \cite{bobb2015bayesian,bobb2018statistical} throughout. 

This non-parametric formulation has the advantage of allowing non-linearity as well as high-order interactions without needing to explicitly include all desired interaction terms {\it a priori}. This flexibility has two drawbacks, however. First, it makes interpretation difficult because one must manually investigate interactions by, say, plotting exposure response curves while holding other exposures at different levels. Second, the model may be too flexible when there are few or no higher order interactions; in such a case BKMR is likely less efficient than a GAM, which assumes additivity among the exposure effects {\it a priori}.

\subsection{MixSelect} \label{ss:mixselect}

\cite{ferrari2020identifying} augmented BKMR by including linear main effects and two-way interactions in an approach called MixSelect. The MixSelect model is 
\begin{align}
y_i|h_i &\sim N(\mathbf{x}_{i}^T\boldsymbol{\beta}+\sum_{j=1}^p \sum_{k>j} {\lambda_{jk}}{x}_{ij}{x}_{ik}+ h_i^*+\mathbf{z}_i^T\boldsymbol{\alpha},\sigma^2). 
\end{align} 
Here $h^*_i$ is the $i^{th}$ element of $[\mathbf{I}-\mathbf{H}]\mathbf{h}$, where $\mathbf{H}$ is the usual hat matrix, i.e. the projection matrix onto the column space of the exposures, and only include the linear main effects. This serves to reduce so-called confounding between the linear main effects and the non-linear $\mathbf{h}$, as in the spatial literature \citep{hanks2015restricted,guan2018computationally}. Then placing separate selection priors on the linear main  and  interactions coefficients $\{\boldsymbol{\beta}, \boldsymbol{\lambda}\}$ and for  non-linear kernel components $\boldsymbol{\rho}$, this allows for non-linear components to be selected out of the model while still allowing linear main and interaction effects. This proved to be more accurate than BKMR in the absence of non-linearity. That being said, if there is in fact are non-linear relationships, this is only captured by the kernel function which is again overly flexible, allowing for high order non additive interactions, which makes interpretation challenging. 

What's more, the $\mathbf{H}$ adjustment does not maintain identifiability due to shared components in the linear two-way interactions and the kernel function. Interestingly, the authors wrote that although one could in theory use the orthogonal projection onto the column space of both linear main and interaction effects, they ``noticed in [their] simulations that this would make the resulting nonparametric surface too restrictive.'' The lack of identifiability makes it is difficult to characterize evidence in favour of non-linearity and non-additivity.

\subsection{BMIMs} \label{ss:bmim}
In exposomic analyses, the number of exposures $p$ is large, and interpreting the fully non-parametric BKMR can be prohibitive. Researchers often have knowledge of different classes of exposures; for example in the HELIX data (see Section~\ref{s:analysis}), 65 exposures are grouped naturally into 13 different groups, including pollutant classes like  pthalates and organochlorines as well as indoor air measurements and traffic density variables. The BMIM  leverages this group structure to reduce the dimensionality of the non-parametric function in BKMR and make more interpretable inferences. 

Suppose the $p$ exposures $\{x_{i1},\cdots,x_{ip}\}$ are partitioned into $M$, $M\in \{1,\dots,p\}$, mutually exclusive groups denoted  $\mathbf{x}_{im}=(x_{im1},\cdots,x_{imL_m})^T$ for $m=1,\dots,M$. The BMIM \citep{mcgee2021bayesian} is
\begin{align*}
y_i &= h(\mathbf{E}_i)+\mathbf{z}_i^T\boldsymbol{\alpha} +\epsilon_i, ~~\epsilon_i\sim N(0,\sigma^2), 
\end{align*}
where $\mathbf{E}_i=(E_{i1},\dots,E_{iM})^T$
is a vector of multi-exposure indices $E_{im}=\mathbf{x}_{im}^T \boldsymbol{\theta}_m$ with  $L_m$-vector of unknown index weights  $\boldsymbol{\theta}_m=(\theta_{m1},\dots,\theta_{mL_m})^T$, where $\theta_{ml}$ quantifies the contribution of the $l^{th}$ component to the $m^{th}$ index.  Note that  $h(\cdot)$ is  now an unknown $M$-dimensional  function of an index vector, whereas in Section \ref{ss:bkmr} it was a $p$-dimensional function of exposures ($M\leq p$). The weights appear in the kernel function, $K(\mathbf{E}_i,\mathbf{E}_{i'})=\exp\left[-\sum_{m=1}^M \rho_m ([\mathbf{x}_{im} -\mathbf{x}_{i'm}]^T \boldsymbol{\theta}_m)^2\right],$ and  need to be constrained for identifiability due to the unknown $\rho_j$: (i) $\boldsymbol{1}_{L_m}^T\boldsymbol{\theta}_m\geq 0$, where $\boldsymbol{1}_{L_m}$ is the unit vector of length $L_m$, and  (ii) $\boldsymbol{\theta}_m^T\boldsymbol{\theta}_m =1$. 
Instead of estimating parameters in this constrained space, we follow \citep{mcgee2021bayesian} and reparameterize the model in terms of unconstrained  $\boldsymbol{\theta}_{m}^*=\rho_m^{1/2}\boldsymbol{\theta}_{m}$, on which one places priors directly. Despite this dimension reduction,  the BMIM suffers from the same drawbacks as BKMR, inextricably linking non-linearity and non-additivity, albeit in lower dimension. 

\section{Proposed Methods} \label{s:ckmr}

\subsection{Collapsible Kernel Machine Regression }\label{ss:ckmr}
We propose a model that addresses these problems by collapsing to an additive structure when suggested by the data, while retaining the flexibility of BKMR when necessary. The collapsible kernel machine regression (CKMR) model is
\begin{align*}
y_i &= \sum_{j=1}^p f_j({x}_{ij})+h^*(\mathbf{x}_i)+\mathbf{z}_i^T\boldsymbol{\alpha} +\epsilon_i, ~~\epsilon_i\sim N(0,\sigma^2), 
\end{align*}
where $f_j(\cdot)$ is a smooth unknown function of the $j^{th}$ exposure, and $h^*(\cdot)$ is a $p$-dimensional non-parametric function of $\mathbf{x}_i$, subject to an identifiability constraint. 

We represent the $f_j(\cdot)$ via basis expansion: $f_j(\cdot)\approx \mathbf{B}^T_{ij} \boldsymbol{\beta}_j$, where $\mathbf{B}_{ij}=[b_{j1}(x_{ij}),\dots b_{jd_j}(x_{ij})]^T$, and $b_{jl}(\cdot)$, $l=1,\dots,d_j$, are known basis functions, and  $\boldsymbol{\beta}_j$ is a $d_j$-vector of coefficients. We use B-splines, though alternative basis functions could be used. We encourage smoothness via a quadratic  roughness penalty   $\frac{1}{2\tau_j^2} \boldsymbol{\beta}_j^T \boldsymbol{S}_j \boldsymbol{\beta}_j$, where $\boldsymbol{S}_j$ is a known $d_j \times d_j$ matrix of second derivatives of basis functions, and $\tau_j^2$ is an unknown smoothing parameter estimated from the data \citep{wood2017p}. 
This is equivalent to (the log of) a mean-zero Gaussian prior with precision $\frac{1}{\tau_j^2}\boldsymbol{S}_j  $.  

To ensure identifiability, $h^*(\mathbf{x}_i)$ must be restricted in some way. We define   $\mathbf{h^*}=[h^*(\mathbf{x}_1),\dots,h^*(\mathbf{x}_n)]^T$  as $\mathbf{h^*}=\boldsymbol{P}\mathbf{h}$, where $\mathbf{h}\sim N(\mathbf{0},\nu^2\sigma^2\mathbf{K} )$ 
as in the mixed effects representation in (\ref{eqn:lme}), $\boldsymbol{P}$ is the projection matrix $\boldsymbol{P}:=\mathbf{I}-\mathbf{B}(\mathbf{B}^T\mathbf{B})^{-1}\mathbf{B}^T$ using a generalized inverse as appropriate, and  $\mathbf{B}$  is the $N\times \sum_{k=1}^p d_k$ design matrix with $i^{th}$ row equal to $\mathbf{B}_i^T$. Equivalently,  $\mathbf{h^*}\sim N(\mathbf{0}, \nu^2\sigma^2 \boldsymbol{P} \mathbf{K}\boldsymbol{P})$. Thus $\mathbf{h^*}$ is the projection of the usual non-parametric component $\mathbf{h}$ onto the orthogonal complement of the column space of $\mathbf{B}$. The columns of $\mathbf{B}$ span the space of smooth \textit{additive} functions of $\mathbf{x}_i$, so $\mathbf{h^*}(\cdot)$ captures \textit{non-additive} deviations.  Inuitively, this decomposes the surface $\mathbf{h}(\cdot)$ into: (i) additive effects, captured by $\sum_{j=1}^p f_j(\cdot)$, and (ii) non-additive interactions, captured by $\mathbf{h^*}(\cdot)$.  We note that although $\mathbf{h^*}(\cdot)$ could technically capture non-linear main effects that are more flexible than allowed by the spline basis, this is negligible when using a rich enough basis; in the penalized splines approach, one generally sets the number of basis function to be fairly large and encourages smoothness via penalization. 

\subsection{Variable Selection \& Hierarchical Constraints} \label{ss:varsel}
We propose a hierarchical variable selection approach that uses two sets of indicators for (i) selection of the main effect of each component and for (ii) selection of each component into the interaction kernel.
To first allow inclusion or exclusion of the $j^{th}$ additive component, we place spike-and-slab priors on the vector $\boldsymbol{\beta}_j$:
\begin{align*}
\boldsymbol{\beta}_j&\sim \gamma_j g(\boldsymbol{\beta}_j)+(1-\gamma_j) \boldsymbol{\delta}_{\mathbf{0}} \\
\gamma_j &\sim \text{Bernoulli}(\pi)  \\
\pi & \sim \text{Beta}(a_{\pi},b_{\pi}),
\end{align*}
where $g(\boldsymbol{\beta}_j)$ is a multivariate zero-mean Gaussian density with precision  $\frac{1}{\tau_j^2}\boldsymbol{S}_j $,  and 
$\boldsymbol{\delta}_{\mathbf{0}}$ represents a point mass at the zero vector. To include or exclude non-additive effects, we place hierarchically constrained spike-and-slab-priors on the non-additive components $\rho_j$. The prior is
\begin{align*}
\rho_j&\sim \gamma_j^{\rho} \text{Gamma}(a^{\rho},b^{\rho})+(1-\gamma_j^{\rho}){\delta}_0 \\
\gamma_j^{\rho}&\sim \gamma_j\text{Bernoulli}(\pi^{\rho})+(1-\gamma_j){\delta}_0  \\
\pi^{\rho} &\sim \text{Beta}(a_{\pi}^{\rho},b_{\pi}^{\rho}). 
\end{align*}
This formulation ensures that if the $j^{th}$ component is included in the kernel (i.e. $\rho_j\neq 0$), then its additive component must also appear ($\gamma_j=1$). 

The hierarchical construction, paired with the model decomposition, has several advantages.First, it allows one to test for both \textit{any} effect of an exposure as well as \textit{non-additive} effects of an exposure using posterior inclusion probabilities. Specifically, $\gamma_j=0$ indicates no effect, $\gamma_j=1$ and $\gamma_j^\rho=0$ indicates additive effect only and $\gamma_j=1$ and $\gamma_j^\rho=1$ indicates a non-additive effect. The last combination $\gamma_j=0$ and $\gamma_j^\rho=1$ is infeasible with the hierarchical structure.  
Second, it simplifies the interpretation task because  one need not investigate all $p(p-1)/2$ two-way exposure response plots when one or more non-additive component are not included in the model; rather, one need only investigate potential two-way interactions for  components jointly selected into the kernel component (i.e. for which $\gamma_j^\rho \gamma_k^\rho=1$) with probability. 
Third, it improves efficiency when there is little evidence of interaction among exposures. In particular, users can tune the prior probability of non-additivity via hyperparameters $\{a_{\pi}^{\rho},b_{\pi}^{\rho}\}$.

\subsection{Adaptive Projection} \label{ss:proj}
The construction $\mathbf{h}^{*}=\mathbf{P}\mathbf{h}$ guarantees identifiability, but we find it overly restrictive in practice. When $p$ is even moderately large, the number of columns of $\mathbf{B}$, $\sum_{j=1}^p d_j$, is  large, and the column space of $\mathbf{B}$ may be rich enough to capture some of the variability  otherwise attributable to non-additive effects. This poses no problem for identifiability, which is addressed by the projection matrix $\mathbf{P}$, but it inhibits interpretation and inference. Consider, for example, a sparse, high dimensional setting in which there is a non-additive interaction between two exposures, and there are many other exposures with no effect at all. If the columns of $\mathbf{B}$ are multi-collinear with an interaction, then, after projection, $\mathbf{h^*}$ only captures a fraction of the variability due to the interaction. As a consequence we may erroneously include additive effects of irrelevant exposures and exclude legitimate non-additive effects of others.

To circumvent this, we propose  an adaptive projection matrix, replacing $\mathbf{P}$ by $\mathbf{P}_{\gamma}:=\mathbf{I}-\mathbf{B}_{\gamma}(\mathbf{B}_{\gamma}^T\mathbf{B}_{\gamma})^{-1}\mathbf{B}_{\gamma}^T$,  where  $\mathbf{B}_{\gamma}$ is constructed by deleting from $\mathbf{B}$ the columns corresponding to components not included in the model.   That is, $\mathbf{B}_{\gamma}$ has $i^{th}$ row equal to $[\mathbf{B}_{i,j_1},\dots, \mathbf{B}_{i,j_D}]$  where $\gamma_{j_1}=\cdots=\gamma_{j_D}=1$ and all other $\gamma_j=0$, $j \notin \{j_1,\dots,j_D \}$. 


\subsection{Collapsible Multiple Index Models} \label{ss:cmim}
We extend the collapsible kernel framework to the multiple index model setting. The proposed collapsible multiple index model (CMIM) is
\begin{align*}
y_i &= \sum_{j=1}^M f_j({E}_{ij})+h^*(\mathbf{E}_i)+\mathbf{z}_i^T\boldsymbol{\alpha} +\epsilon_i, ~~\epsilon_i\sim N(0,\sigma^2), 
\end{align*}
where $h^*(\cdot)$ is  an $M$-dimensional non-parametric function of a multi-exposure index vector subject to an identifiability constraint, and $f_j(\cdot)$ is a smooth unknown function of the $j^{th}$ index. 

We again approximate the $f_j(\cdot)$ via basis expansions, $f_j(\cdot)\approx \mathbf{B}^T_{ij} \boldsymbol{\beta}_j$, where $\mathbf{B}_{ij}=[b_{j1}(E_{ij}),\dots, b_{jd_j}(E_{ij})]^T$, $\boldsymbol{\beta}_j$ is a $d_j$-vector of coefficients and the $b_{jl}(\cdot)$, $l=1,\dots,d_j$, are known basis functions. However, a key difference and complicating factor for the CMIM is that these basis functions are now functions of unknown weights $\boldsymbol{\theta}_m$. We impose smoothness via the same quadratic penalty as before, which is equivalent to adopting a mean-zero Gaussian prior with precision $\frac{1}{\tau_j^2}\boldsymbol{S}_j$.

Unlike the previous framework, $\mathbf{B}_{ij}$ now depends on unknown parameters $\boldsymbol{\theta}_m$; hence, we must update $\mathbf{B}_{ij}$ within the MCMC. The penalty matrices $\boldsymbol{S}_j$ are no more complicated than before, since they are  functions of the known basis functions and not the underlying indices, but we need to update $\mathbf{B}_{ij}$ for every new value for $E_{ij}=\mathbf{x}_{ij}^T \boldsymbol{\theta}_j$ (i.e. whenever $\boldsymbol{\theta}_j$ is updated).   This is not computationally intensive, as the basis functions $b_{jl}(\cdot)$ are known and can be easily evaluated at any new value for $E_{ij}=\mathbf{x}_{ij}^T \boldsymbol{\theta}_j$, for fixed knots. 
Second, $\boldsymbol{\theta}_j$ now appears in both the non-additive \textit{and} the additive components of the model, and both components need to be included in Metropolis-Hastings steps.

Reparameterizing the kernel in terms of unconstrained $\boldsymbol{\theta}^*_m$ as in the standard BMIM \citep{mcgee2021bayesian,wilson2022kernel} is no longer possible because the basis functions $b_{jd}(\cdot)$ are non-linear functions of $E_{ij}=\mathbf{x}_{ij}^T \boldsymbol{\theta}_j$. Instead we build on the approach of \cite{antoniadis2004bayesian}, which used Fisher von-Mises priors (and proposals) for weights in single index models. We augment the hierarchical prior structure in Section \ref{ss:varsel} to be
\begin{align*}
\boldsymbol{\theta}_j &\sim \gamma_j \text{vonMisesFisher}(\kappa,\boldsymbol{\mu}) + (1-\gamma_j) \boldsymbol{\delta}_{\boldsymbol{\mu}_{L_m}},
\end{align*}
where $\boldsymbol{\delta}_{\boldsymbol{\mu}_{L_m}}$ is a point mass at $L_m^{-1/2}\mathbf{1}_{L_m}$, and $\kappa$ and $\boldsymbol{\mu}$ are hyperparameters representing the concentration and the mean direction of the distribution.  Absent prior knowledge of relative weights, we set $\boldsymbol{\mu}=L_m^{-1/2}\mathbf{1}_{L_m}$.

We again employ the adaptive projection approach of Section \ref{ss:proj}. Paired with the hierarchical variable selection, the proposed model collapses to an additive index structure where possible while still allowing for non-additive interactions among indices where appropriate.

Posterior inference follows via MCMC sampling; see supplementary Appendix A.2 for complete details on the sampler. In particular, we implement the Gaussian predictive process approximation of \cite{bobb2018statistical} and \cite{savitsky2011variable} to facilitate faster computation when $N$ is large (see supplementary Appendix A.3).

\subsection{Extensions}
Extending to the proposed collapsible framework to BMIMs yields several convenient models. First, it extends naturally to identifying  windows of susceptibility given time-varying exposures. Suppose $M$ exposures are measured longitudinally, and $\mathbf{x}_{im}=(x_{im1},\cdots,x_{imT})^T$ corresponds to the $m^{th}$ exposure measured at $T$ times. Following  \cite{wilson2022kernel}, we take a functional approach and represent $\mathbf{x}_{im}$ and weights $\boldsymbol{\theta}_m$ using the same orthonormal basis: $\boldsymbol{\theta}_m=\mathbf{\Psi}_m \tilde{\boldsymbol{\theta}}_m$ and  $\mathbf{x}_{im}=\mathbf{\Psi}_m\boldsymbol{\xi}_{im}$, where $\mathbf{\Psi}_m$ is a matrix whose columns define an orthogonal basis expansion, and $\tilde{\boldsymbol{\theta}}_m$ and $\boldsymbol{\xi}_{im}$ are unknown coefficients. A least squares approximation then yields $E_{im}=\mathbf{x}_{im}^T \mathbf{\Psi}_m \tilde{\boldsymbol{\theta}}_m={\tilde{\mathbf{x}}_{im}}^T \tilde{\boldsymbol{\theta}}_m$,  which takes the form of a linear index with transformed exposure vector $\tilde{\mathbf{x}}_{im}=\mathbf{\Psi}_m^T \mathbf{x}_{im}$. The upshot is that we can apply a known linear transformation $\mathbf{\Psi}_m^T$ to $\mathbf{x}_{i}$ and place priors directly on $\tilde{\boldsymbol{\theta}}_m$. Estimation then follows as above, and posterior draws of $\boldsymbol{\theta}_m$---indicating times at which exposure effects are strongest---are recovered by transforming posterior draws of $\tilde{\boldsymbol{\theta}}_m$.

Second, a user can incorporate prior knowledge about the epxosures into the index structure in a way that would be challenging in a fully non-parametric approach \citep{mcgee2022integrating} . For example, when exposures are believed to act in the same direction (i.e., either all  protective or all detrimental), then one can incorporate this information to improve efficiency by adopting a non-informative Dirichlet prior on index weights $\boldsymbol{\theta}_m$ (under the slightly different constraint that $\mathbf{1}^T\boldsymbol{\theta}_m$=1). When one further has knowledge about the relative magnitudes of exposure effects, for example based on toxic equivalency factors from toxicology research, one can incorporate this via informative Dirichlet priors or hard constraints on index weight orderings (see \citealp{mcgee2022integrating} for details).

\section{Simulations} \label{s:sims}

We conducted a series of simulations to investigate the performance of the proposed methods. The goals of the simulations were to: (i) compare accuracy and efficiency relative to existing methods; (ii) verify that the proposed approaches yield valid inference; and (iii) show that the proposed methods can quantify evidence of non-additive interactions.  We considered both standard (non-index) models as well as multiple index models, and we investigated two scenarios, corresponding to: (A) no interactions and (B) interactions between some exposures/indices.

\subsection{Setup}\label{ss:setup}
We consider 4 scenarios. For each scenario we generated $R=500$ data sets of $N$=$200$ observations. Scenarios A and B are non-index scenarios, we first drew ten standardized uniformly-distributed exposures, $\{x_1,\dots,x_{10}\}$, as well as two independent continuous covariates, $\{z_1,z_2\}$, treated linearly.  We then generated outcomes according to one of two exposure-response functions. In both, six of the ten exposures had non-null effects, four of which were non-linear. In Scenario A, there were no interactions. We generated outcomes as $y=\mu_A +0.5 z_1+\epsilon$, where $$\mu_A=2cos(2 x_1) + x_2 +4 f_{t}(2x_3) +sin(2x_4)+x_5^2 -x_6,$$ 
$f_t(\dot)$ is the density function of the student's t-distribution with ten degrees of freedom, and $\epsilon \sim N(0,\sigma^2)$. In Scenario B, we included an interaction between $x_1$ and $x_5$. We generated outcomes as above, replacing $\mu_A$  with $\mu_B=\mu_A + cos(2 x_1)x_5^2$.  In the main simulations we set $\sigma^2=1$, and additionally considered a higher noise setting in the supplementary material where $\sigma^2=2$.

Scenarios C and D are multiple index scenarios. We replaced the first two exposures with multipollutant indices containing four exposures each. We first generated sixteen exposures $\{x_1,\dots,x_{16}\}$. We then created two indices each consisting of four components. Those indices are $E_1=3 x_1 +2 x_{11} +1 x_{12 }+0 x_{13}$ and $E_2=x_2+x_{14}+x_{15}+x_{16}$. The remainder of the exposures are each in their own index, $E_j=x_j$ for $j=3,\dots,10$. Thus we generated from a 10-index model with $(L_1,\dots,L_{10})^T=(4,4,1,\dots,1)^T$. Scenario C is the same as scenario $A$ expect the exposure-response function is a function of the indices $E_1,\dots,E_{10}$. Scenario C is the same as Scenario B with index exposures as the input and included an interaction between   multi-exposure index $E_1$ and single component index $E_5=x_5$.

In each dataset we estimated the unknown exposure-response surface with several methods. In the non-index setting, we fit the proposed CKMR (with 9 spline degrees of freedom), standard BKMR (via the {\tt bkmr} package; \citep{bobb2018statistical}), and two competing methods. As a gold standard when there are no interactions, we fit GAM with spike-and-slab priors (ssGAM) via the {\tt spikeslabGAM} package in {\tt R} \citep{scheipl2012spike}. 
We also fit the non-linear interaction  (NLinter) approach of \cite{antonelli2020estimating}, which allows for separate selection of additive and non-additive effects via natural cubic splines with sparsity inducing priors (and we used 6 spline degrees of freedom), and the  MixSelect approach of \cite{ferrari2020identifying}.  In the index setting, we fit the proposed CMIM and the standard BMIM. We also fit a version of CKMR \& CMIM that does not use the adaptive projection approach of \ref{ss:proj} (that is, $\mathbf{P}$ is computed once as the orthogonal projection matrix for the column space of all basis functions regardless of whether some main effects are selected out of the model).  

To quantify accuracy and efficiency, we report mean squared error (MSE) and 95\% credible interval width for the true exposure-response surface, $h(\cdot)$, averaged over observed exposure levels. To assess validity we then report bias and interval coverage. To show the proposed methods can identify evidence of non-additivity, we compute average posterior inclusion probabilities (PIPs) of main effects as well as of interactions (i.e. inclusion in the kernel). In particular, we report joint posterior probabilities that $\gamma^\rho_j \gamma^\rho_k =1$, indicating joint selection into the kernel and thus potential pairwise interaction. Other than these proposed interaction PIPs, only NLinter offers PIPs for interactions.


\subsection{Results}\label{ss:simresults}
We report average MSE and interval width in Table \ref{tab:simtab} (and boxplots can be found in the supplementary material). Consider first the non-index setting. 
	When there was no interaction (Scenario A), BKMR performed worst, with an  average MSE over 50\% higher than all others, as it did not exploit the lack of interactions. All other methods yielded comparable MSE and good coverage; in particular, CKMR performed nearly as well as the GAM, which would be the optimal approach in this scenario.
	When there was interaction (Scenario B),  ssGAM naturally yielded very bad coverage, because it erroneously assumed additivity. By contrast, the proposed CKMR performed best, yielding the lowest MSE and nominal coverage.
	Though we expected the greatest gains in the absence of interactions,  BKMR was still outperformed by CKMR in Scenario B. 
	This is because even though there is indeed an important interaction between two exposures, the other exposures did not interact; CKMR was able to exploit this where BKMR could not.

\begin{table}[htbp!]
	\caption{Simulation results  with $\sigma^2=1$: comparing proposed CKMR to BKMR, a non-linear interaction selection method (NLInter), and a spike-and-slab GAM (ssGAM) when data are generated from a componentwise model with and without interaction. And comparing proposed CMIM to BMIM when data are generated from an indexwise model with and without interaction.} \label{tab:simtab}
	\centering
	\begin{tabular}{llllrrrr}
		\toprule
	Scenario &	Indexwise & Interaction & Method & MSE & Bias & Width & Cvg  \\ 
		\midrule
	A	&No & No 		&  ssGAM & 0.35 & -0.01 & 1.62 & 0.98 \\ 
		&&   & BKMR & 0.67 & -0.01 & 2.22 & 0.89 \\ 
		&&    &    NLInter & 0.44 & -0.00 & 1.68 & 0.94 \\ 
		&&    &    NonAdaptive & 0.38 & -0.01 & 1.60 & 0.97 \\ 
		&&    &    CKMR & 0.38 & -0.01 & 1.60 & 0.96 \\ 
		\midrule
		B &No& Yes 		&  ssGAM & 0.69 & -0.01 & 1.91 & 0.84 \\ 
		&&    & BKMR & 0.72 & -0.00 & 2.36 & 0.88 \\ 
		&& &     NLInter & 0.69 & -0.00 & 2.09 & 0.87 \\ 
		&& &     NonAdaptive & 0.53 & -0.01 & 1.97 & 0.94 \\ 
		&& &     CKMR & 0.49 & -0.01 & 1.87 & 0.95 \\ 
		
		\midrule
		C &Yes & No & BMIM & 0.68 & -0.01 & 2.46 & 0.92 \\ 
		&&    & NonAdaptive & 0.44 & -0.01 & 1.77 & 0.95 \\ 
		&&    & CMIM & 0.43 & -0.01 & 1.76 & 0.96 \\ 
	\midrule	
		D &Yes& Yes & BMIM & 0.73 & -0.01 & 2.56 & 0.91 \\ 
		&&     & NonAdaptive & 0.65 & -0.01 & 2.18 & 0.91 \\ 
		&&     & CMIM & 0.56 & -0.01 & 2.08 & 0.94 \\ 
		\bottomrule
	\end{tabular}
\end{table}

We found similar results in the multiple index setting. When there was no interaction (Scenario A), BMIM was unable to exploit the absence of interactions, and yielded 50\% higher MSE and 40\% wider intervals than CMIM. When there was an interaction (Scenario B), CMIM still performed better than BMIM but the difference was smaller (BMIM had 30\% higher MSE), again because CMIM exploited the fact that not all exposures had interactions.

\begin{landscape}
	\begin{figure}[htbp!]
		\centering
		\includegraphics[width=0.49\linewidth]{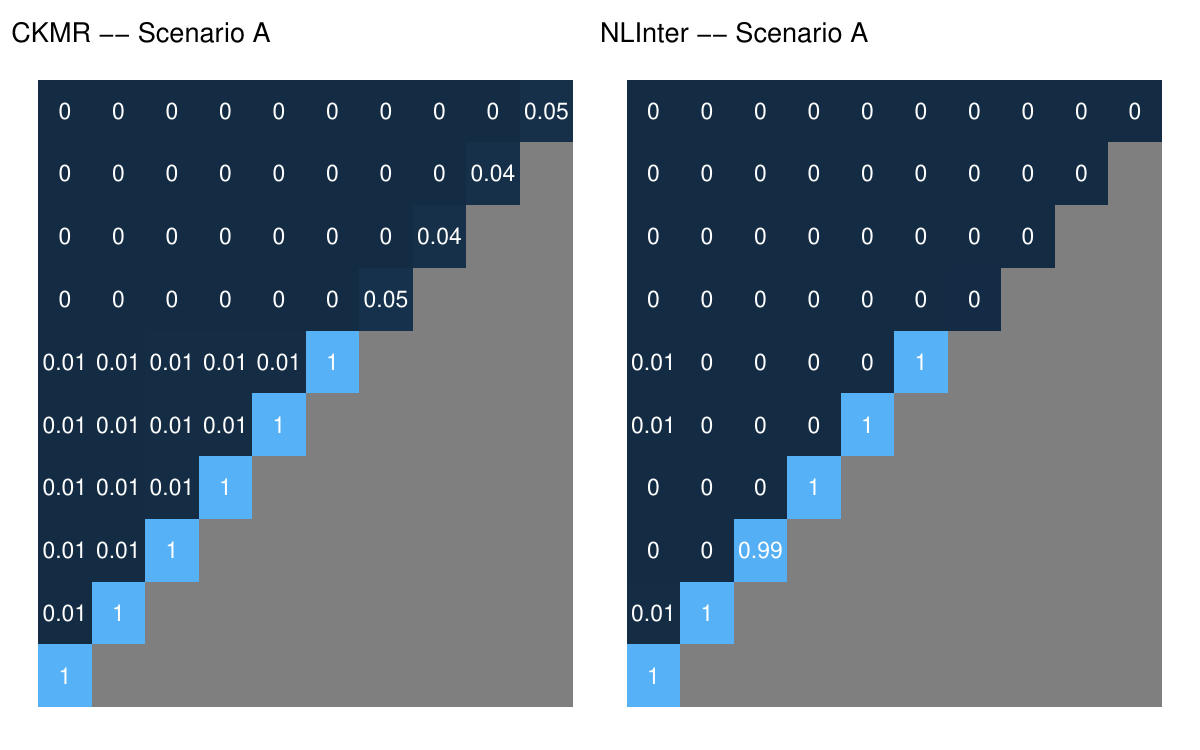}~~~~\quad
		\includegraphics[width=0.49\linewidth]{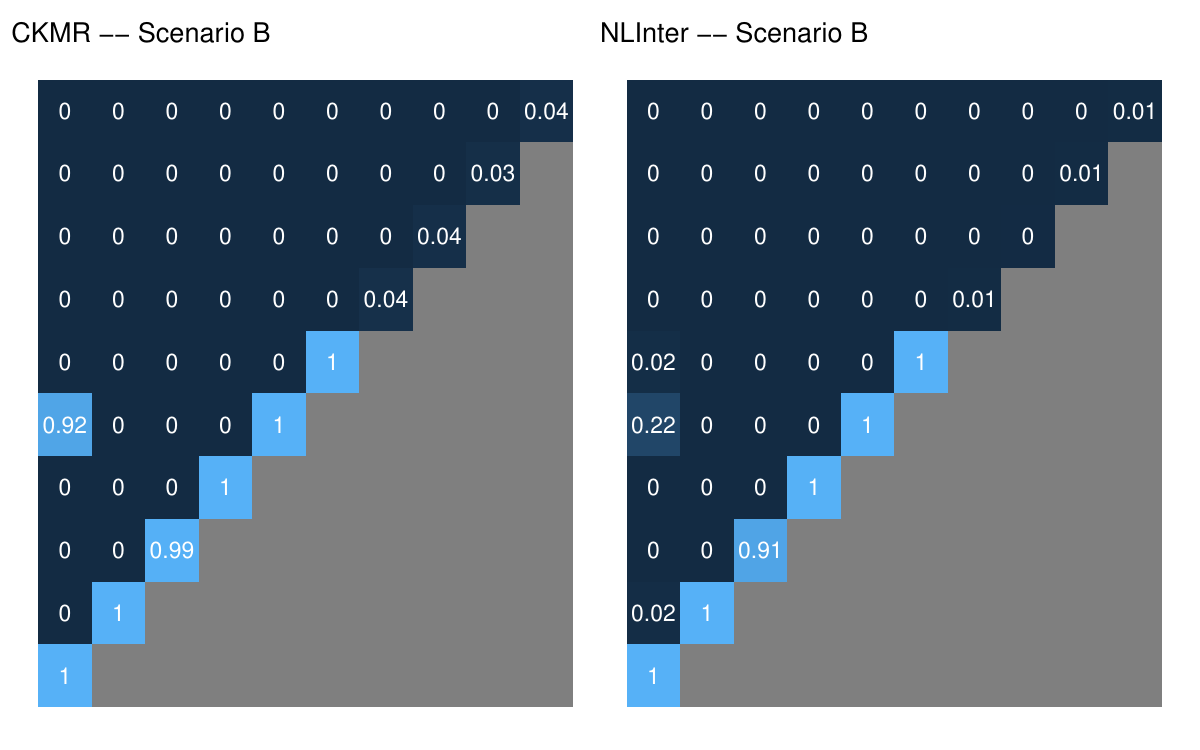} \\
		\includegraphics[width=0.49\linewidth]{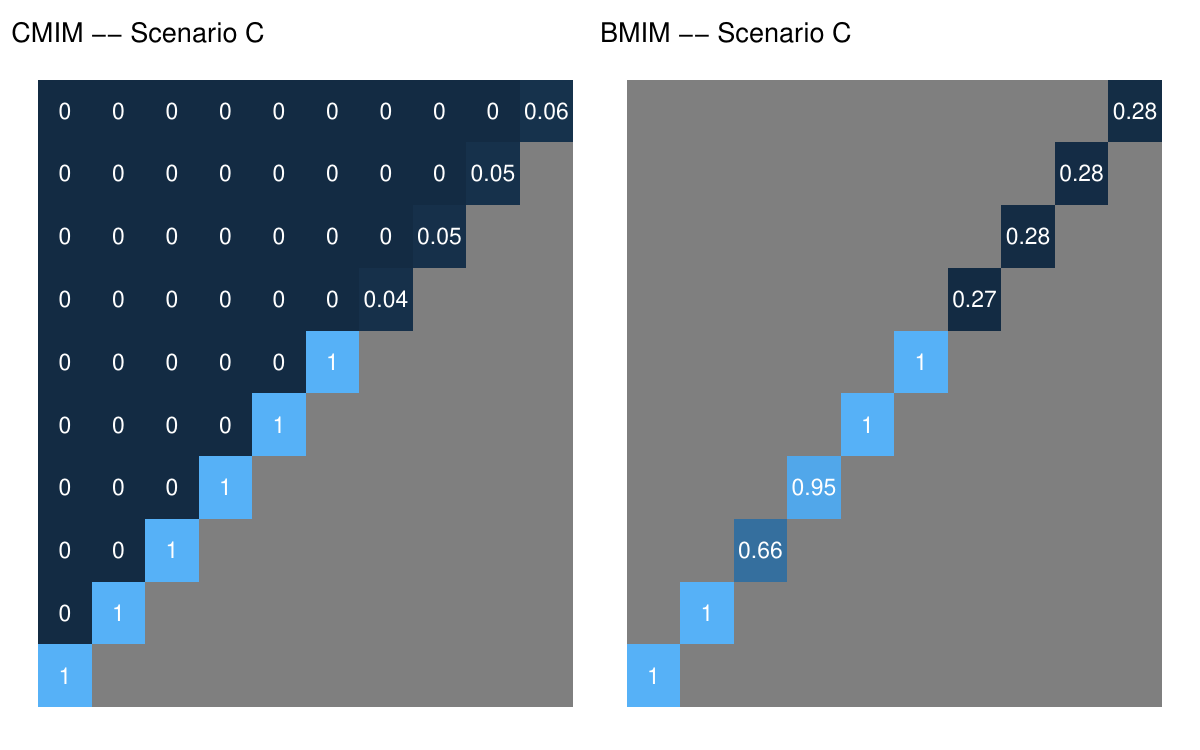}~~~~\quad
		\includegraphics[width=0.49\linewidth]{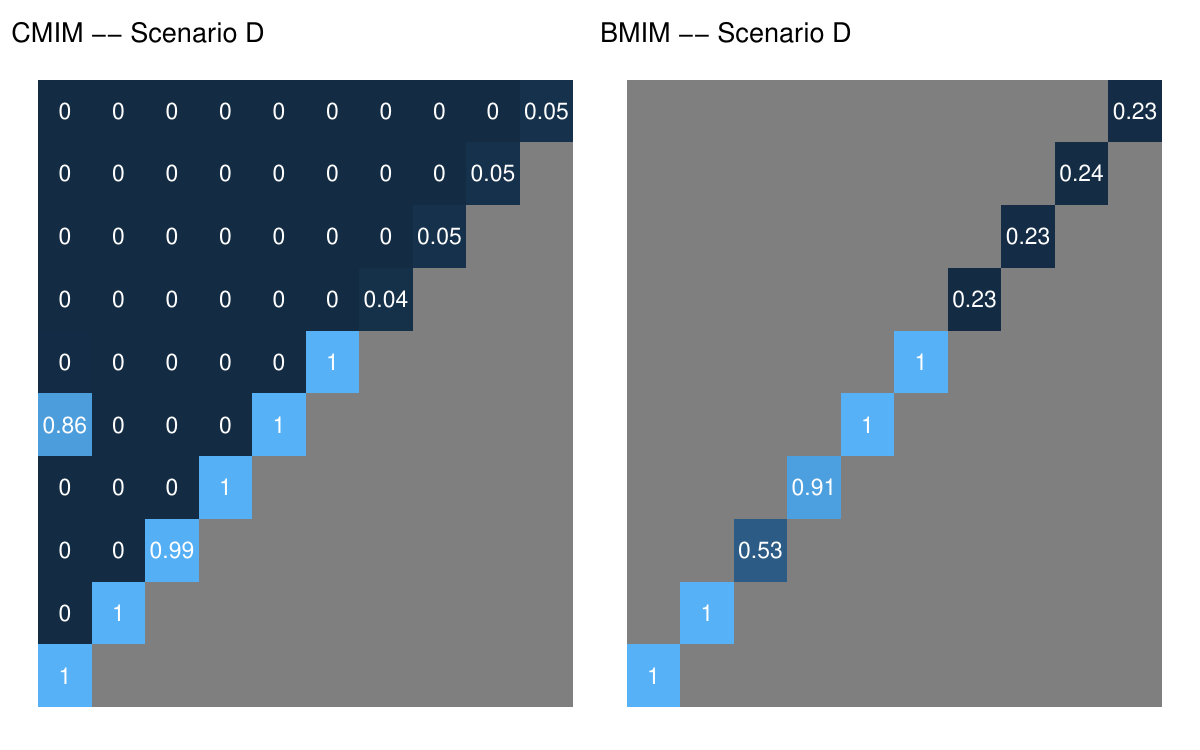}
		\caption{Heatplot of main+Interaction PIPs in componentwise (top row) and indexwise (bottom row) simulations with $\sigma^2=1$. First 6 components are non-null. In non-additive simulations (right two columns), components 1 \& 5 interact.}
		\label{fig:PIPs}
	\end{figure}
\end{landscape}

We plot average main effect and interaction PIPs in Figure \ref{fig:PIPs}.  For the componentwise setting, all methods assigned high posterior probability to the six non-null effects and low probability to the four null effects most of the time. Only CKMR  and NLInter provide PIPs for interactions, and we we see that the proposed approach has much higher power to detect an interaction than NLInter (92\% vs 22\%). In the higher noise setting we found similar results but lower power overall: (41\% vs 8\%; see Supplementary Material). In the multiple-index models, the BMIM had lower power to detect the $x_3$ main effect, with PIPs of 66\% and 53\% with and without interaction, respectively, and had slightly elevated PIPs for null effects. By contrast the proposed CMIM correctly captured all non-null main and interaction effects with high PIPs, and yielded low PIPs for null effects. 

Moreover we find 
that unlike the proposed approach, the non-adaptive version failed to accurately pick up interactions, with average PIPs of 69\% rather than 92\% (and 24\% rather than 41\% in the high noise setting) in the componentwise setting, with similar results in the indexwise setting (see Figures \ref{fig:suppPIPsNonadaptive} and \ref{fig:suppPIPsNonadaptive2}). This led to higher MSEs when data were generated with interactions: e.g., 0.65 vs 0.56 in scenario D.

\section{Case Study: HELIX} \label{s:analysis}

\subsection{Setup} 
We analyze data from the Human Early Life Exposome (HELIX) project \citep{maitre2018human,vrijheid2014human}, a large study of environmental and chemical exposures among of 30,000 mother-child pairs in six European countries. In particular, we are interested in early life (postnatal) exposures and their relationship with childhood BMI, so we analyzed data from a subcohort of children for which childhood BMI was measured, made available by The Barcelona Institute for Global Health (ISGlobal). To protect participant privacy while making data publicly available, ISGlobal used a combination of real and simulated data; details on the data anonymization process can be found in \cite{maitre2022state}. Data can be downloaded at \href{github/isglobal-exposomeHub}{https://github.com/isglobal-exposomeHub/ExposomeDataChallenge2021}.  

\begin{figure}[htbp!]
	\centering
	\includegraphics[width=0.99\linewidth]{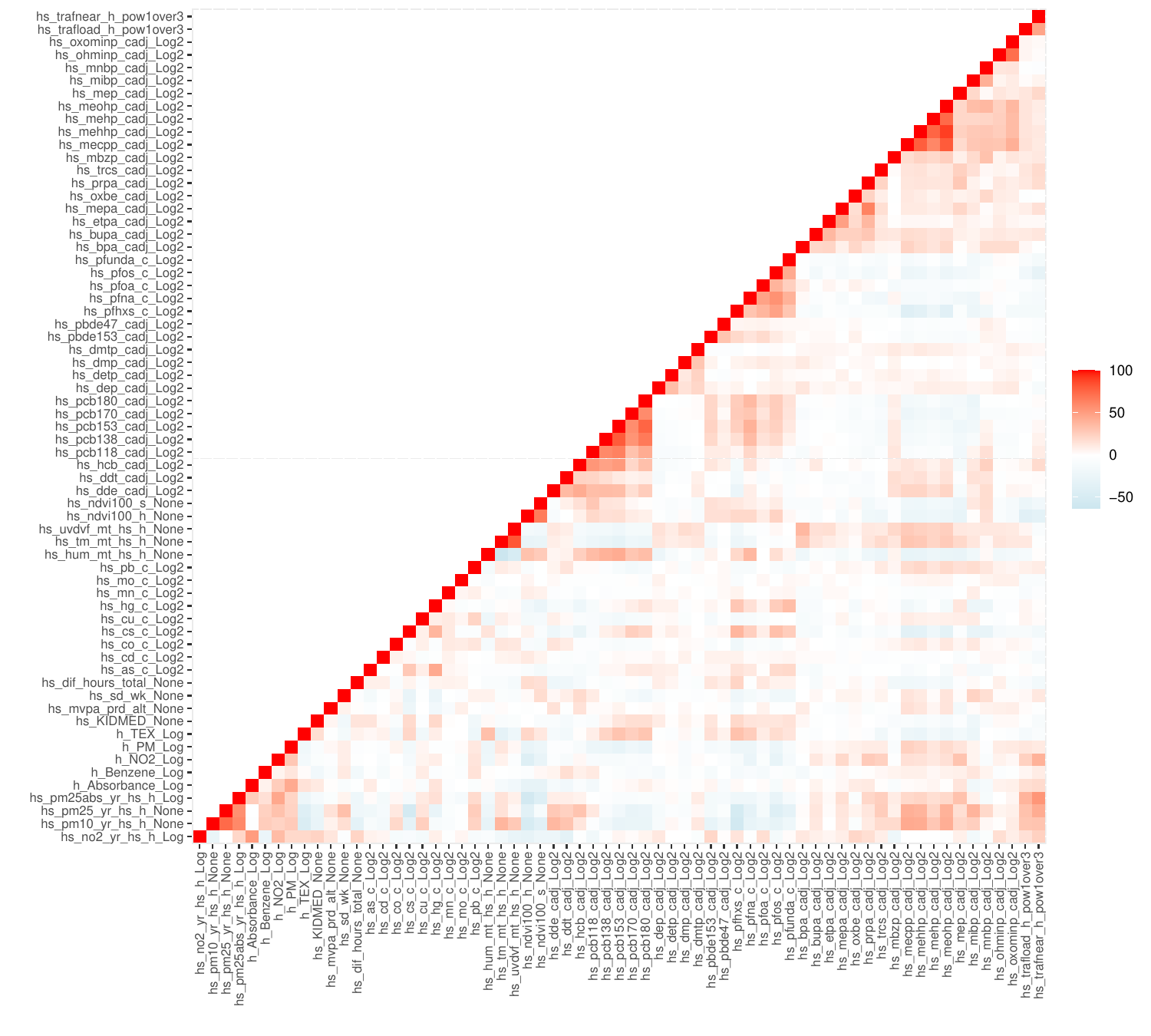}
	\caption{Heatplot of correlations among mixture components in HELIX analysis.}
	\label{fig:corrplot}
\end{figure}

We observed age- and sex-standardized BMI for $N$=1,301 children aged 6 to 11. For each subject we also observed $p$=65 postnatal exposures, which were naturally pre-grouped into 13 classes: air pollution variables, indoor air measurements, lifestyle factors, metals, meteorological variables, natural spaces, organochlorines, organophosphates, PBDEs, PFAS, phenols, phthalates, and traffic density variables. Exposures were strongly correlated within classes, with correlations as high as 0.9 among phthalates and 0.8 among organochlorines (see Figure \ref{fig:corrplot}). Cross correlations were lower between different classes, with a maximum of 0.5. Due to the high dimensionality, the strong within-class correlations, and expected similarity behaviour of exposures within a class, we used these classes as index groupings and thus fit a $M$=13-index model via the proposed CMIM approach in addition to the existing BMIM approach. These allowed for 13 distinct non-linear indexwise curves, in which interactions  were permitted among different indices but not within indices. That is, metals and phenols may interact, but two different metals may not. We also adjusted (linearly) for several maternal covariates consisting of maternal age, pre-pregnancy BMI, pregnancy weight gain, education level, parity before pregnancy, and cohort, as well as several child-level covariates consisting of gestational age, year of birth, and number of parents native to the country.

\subsection{Results} 
\begin{figure}[hbp!]
	\centering
	\includegraphics[width=0.99\linewidth]{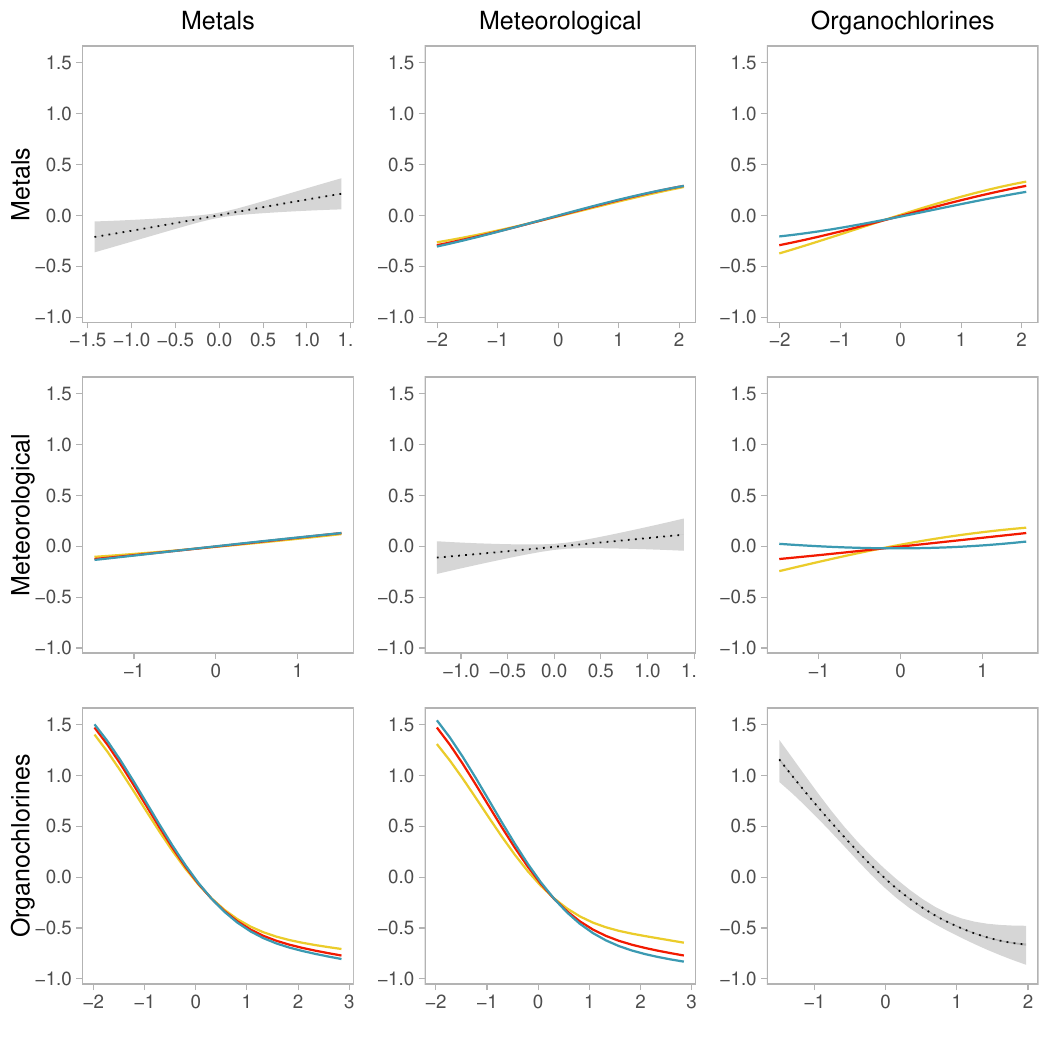} 
	\caption{HELIX analysis: results of standard BMIM. Plots on diagonal are indexwise curves with 95\% credible intervals, with other indices set to their medians. Off diagonals are two-way interaction plots, setting second index (along x-axis) to 10th, 50th and 90th percentiles. }
	\label{fig:bmim_inter}
\end{figure}

The BMIM yielded high indexwise PIPs for metals, meteorological variables and organochlorines (see Table \ref{tab:PIPs}). Estimated indexwise curves (when holding all other exposures at their medians) indicated moderate linear associations with metals and meteorological variables, and a strong non-linear association with organochlorines (see indexwise curves in Figure \ref{fig:bmim_inter}). Further investigation of two-way interaction plots---plotting indexwise curves while holding another index at 10th, 50th and 90th percentiles---suggests some indication of modest interaction between meteorological variables and organochlorines, but it is difficult to tell given the uncertainty. 

 \begin{table}[htbp!]
 \caption{HELIX Analysis: Index class, size ($L_m$) and PIPs from CMIM.}\label{tab:PIPs}
\centering
\begin{tabular}{lrrr}
  \hline
  && \multicolumn{2}{c}{PIP} \\
		\cmidrule{3-4}
  Index & $L_m$ & BMIM& CMIM \\ 
  \hline
 AirPollution &   4 &33& 22 \\ 
 Indoorair &   5 &36& 17 \\ 
 Lifestyle &   4 &42& 3 \\ 
 Metals &   9 &100& 100 \\ 
 Meteorological &   3 &82& 5 \\ 
 NaturalSpaces &   2 &4& 0 \\ 
 Organochlorines &   8 &100& 100 \\ 
 Organophosphates &   4 &11& 2 \\ 
 PBDE &   2 &34& 7 \\ 
 PFAS &   5 &20& 2 \\ 
 Phenols &   7 &26& 3 \\ 
 Phthalates &  10 &46& 3 \\ 
 Traffic &   2 &8& 0 \\ 
   \hline
\end{tabular}
\end{table}

The CMIM only yielded high PIPs for metals and organochlorines; all others had PIPs below 25\%. Moreover, the CMIM approach quantifies evidence in favour of interaction, and we found no evidence of interaction (two-way interaction PIPs near 0). This allows us to focus the interpretation task on the main effects of the non-null indices.

\begin{figure}[htbp!]
	\centering
	\includegraphics[width=0.45\linewidth]{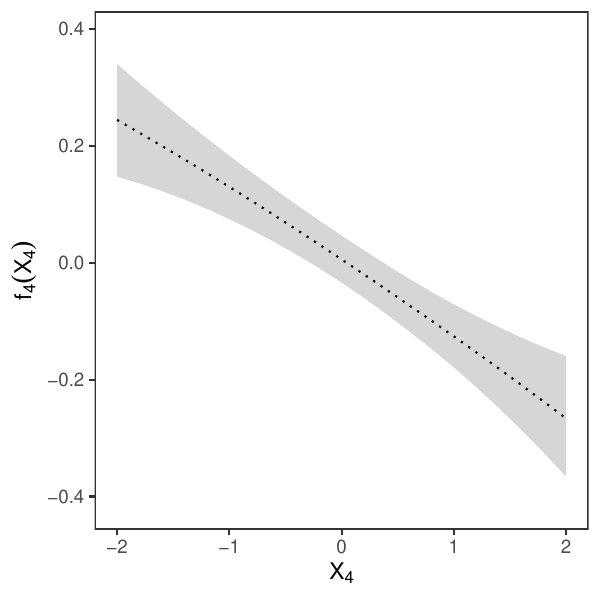}
	\includegraphics[width=0.45\linewidth]{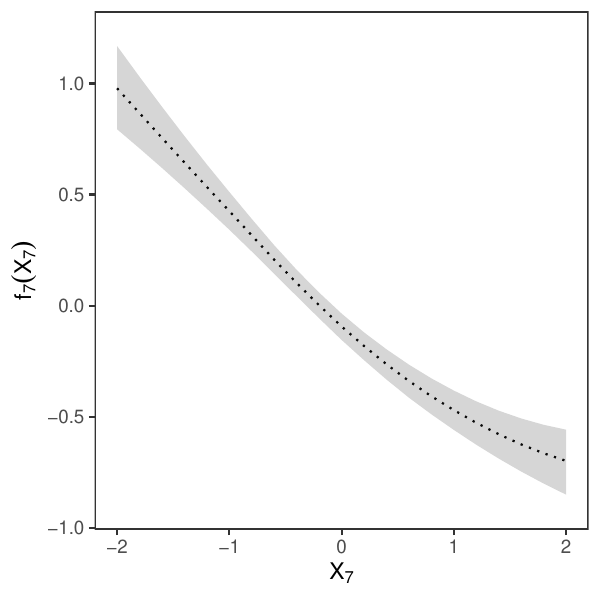}  \\
	\includegraphics[width=0.97\linewidth]{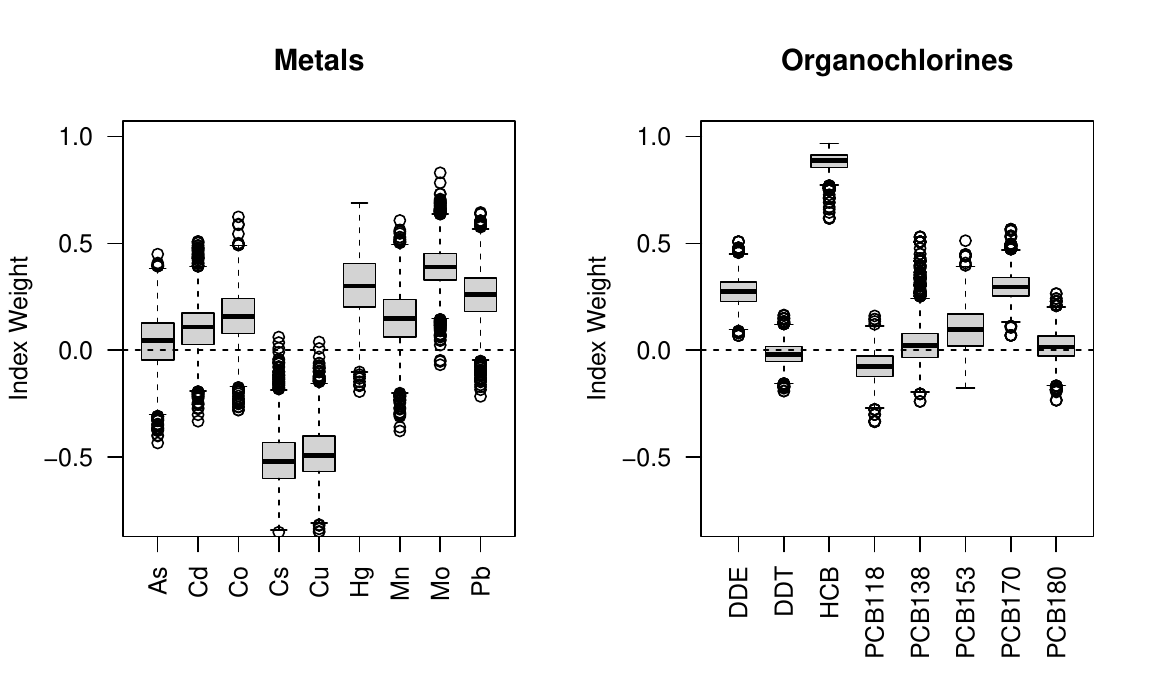} 
	\caption{HELIX analysis results. Top row: estimated indexwise exposure-response functions for metals (left) and organochlorines (right). Bottom row: boxplots of index weights posteriors for metals (left) and organochlorines (right).}
	\label{fig:HELIX}
\end{figure}

We plot the estimated indexwise exposure-response curves from the CMIM in the top row of Figure \ref{fig:HELIX}. Metals had a moderate linear association with BMI; we note that the association appears negative whereas it looked positive in the BMIM approach, but closer investigation of the index weights indicate that the underlying component-wise associations remain in the same direction.  Organochlorines had a stronger non-linear association, with dramatic reductions in mean BMI at low levels of the index, and smaller reductions at higher levels of the index. 

In addition to reducing the dimension of the problem, the index modelling strategy allows us to quantify the contribution of each exposure component to the index. In Figure \ref{fig:HELIX} (bottom row), we plot the posterior distributions for the index weights $\theta_{ml}$. Among metals, the strongest contributions were from cesium (posterior median -0.52; 95\% credible interval [-0.73, -0.24]) and copper (-0.49; 95\% CI [-0.70, -0.22]); in particular they had negative index weights, indicating positive linear associations with BMI (because the slope of the indexwise curve is also negative). Molybdenum and lead both had smaller but non-null associations in the opposite direction (0.39 95\% CI [0.18, 0.61]  and 0.26 95\% CI [0.00, 0.49]).  Among organochlorines, HCB (hexachlorobenzene) was the dominant mixture component (0.89; 95\%CI [0.78, 0.94]), and DDE (dichlorodiphenyldichloroethylene) and PCB170 (0.30; 95\%CI [0.18, 0.45]) also had significant contributions. All other organochlorines had small weights and their intervals covered zero.

\section{Discussion} \label{s:disc}
The proposed CKMR approach is a scalable method for exposome health studies. The work extends the popular BKMR method to the exposure by adding efficiency and interpretability. In addition, we extend the approach to multiple index models, and successfully applied it to an exposome analysis, in which $p=65$ exposures are grouped into $13$ classes.

This work is related to previous projection schemes in the spatial \citep{hanks2015restricted,guan2018computationally} and environmental literature \citep{ferrari2020identifying}. \cite{ferrari2020identifying} modelled linear main effects and two-way interaction effects outside of a kernel component and projected out the linear main effects to reduce ``confounding'' between the kernel component and the main effects. By splitting the linear from the non-linear, this improved efficiency when there were few non linear effects. In contrast, our approach splits the non-linear from the non-additive, which improves efficiency when there is a lack of interactions. Moreover, it streamlines the interpretation task, as one need not investigate all $p(p-1)/2$ two-way interaction plots. Interestingly, \cite{ferrari2020identifying} did not enforce identifiability, projecting out only the linear effects and not the interactions, because doing so would overly restrict the kernel component. A goal of our work is to quantify evidence of non-additive interaction, necessitating identifiability. We found that using a full projection matrix led to poor performance,  exacerbated perhaps by the large design matrix $\mathbf{B}$ whose $\sum_j^p d_j$ columns span the  rich space of smooth additive functions. We found that the issue was due to multicollinearity between the additive effects and non-additive interactions and circumvented it via a novel adaptive projection scheme that maintined identifiability and correctly distinguished non-linear main and interaction effects.

\section{Code and Data Availability}
{\tt R} code  to replicate simulations and data analysis will be available on {\tt Github} upon publication.

The data were created as part of the ISGlobal Exposome data challenge 2021, presented in this publication \citep{maitre2022state}. The HELIX study \citep{vrijheid2014human,maitre2018human}   represents a collaborative project across six established and ongoing longitudinal population-based birth cohort studies in six European countries (France, Greece, Lithuania, Norway, Spain, and the United Kingdom). The research leading to these results has received funding from the European Community’s Seventh Framework Programme (FP7/2007-2013) under grant agreement no 308333 --- the HELIX project and the H2020-EU.3.1.2. --- Preventing Disease Programme under grant agreement no 874583 (ATHLETE project). The data used for the analyses described in this manuscript were obtained from: \href{https://figshare.com/account/home#/projects/98813}{Figshare} (project number 98813) and \href{https://github.com/isglobal-exposomeHub/ExposomeDataChallenge2021/}{Github}  (accessed on 09/10/2024).

\section{Acknowledgement}
We acknowledge the support of the Natural Sciences and Engineering Research Council of Canada (NSERC), DGECR-2022-00433 and RGPIN-2022-03068. This research was also supported by NIH grants ES000002,ES030990 and ES035735.

\bibliographystyle{apalike}
\bibliography{McGee_Bibliography}

\clearpage
\section*{Supplementary Material for ``Collapsible Kernel Machine Regression for Exposomic Analyses'' }
\renewcommand{\thesection}{\Alph{section}}
\setcounter{section}{0}

\renewcommand{\thetable}{\Alph{section}.\arabic{table}}
\renewcommand{\thefigure}{\Alph{section}.\arabic{figure}}

\section{Technical Details}

\subsection{Full Hierarchical Specification} 

For simplicity we let $\mathbf{B}_i=[\mathbf{B}_{i1}^T,\dots,\mathbf{B}_{ip}^T]^T$ and $\boldsymbol{\beta}=[\boldsymbol{\beta}_1^T,\dots,\boldsymbol{\beta}_p^T]^T$, and we rewrite $\sum_{j=1}^p f_j({x}_{ij}) \approx \mathbf{B}_i^T \boldsymbol{\beta}$, subject to the quadratic penalty $\frac{1}{2}\boldsymbol{\beta}^T(\sum_{j=1}^p \frac{1}{\tau_j^2}\boldsymbol{S^*}_j)\boldsymbol{\beta}$, where $\boldsymbol{S^*}_j$ is the expanded  block diagonal matrix with $j^{th}$ block $\boldsymbol{S}_j$ and all other elements zero. 
Also let $\mathbf{B}_{\theta}=[\mathbf{B}_1^T,\dots,\mathbf{B}_N^T]$, where the indicates that this may be a function of index weights $\boldsymbol{\theta}$ in the CMIM setting. 

\begin{alignat*}{2}
	\text{Likelihood}
	&\begin{cases}
		\quad\begin{array}{r@{\hspace{3pt}}l}
			\mathbf{y}|\mathbf{h} &\sim {N}(\mathbf{B}_{\boldsymbol{\theta}}\boldsymbol{\beta}+\boldsymbol{P}_{\boldsymbol{\theta}}\mathbf{h}+\mathbf{Z}\boldsymbol{\alpha},\sigma^2), \text{ where }  \boldsymbol{P}_{\boldsymbol{\theta}}=\mathbf{I}-\mathbf{B}_{\boldsymbol{\theta}}(\mathbf{B}_{\boldsymbol{\theta}}^T\mathbf{B}_{\boldsymbol{\theta}})^{-1}\mathbf{B}_{\boldsymbol{\theta}}^T \\
			\mathbf{h} &\sim {N}(\mathbf{0},\nu^2\sigma^2  \mathbf{K}) \text{ where }\mathbf{K}_{ij}=K(\mathbf{E_i},\mathbf{E_j}) \\
			\implies \mathbf{y}&\sim {N}(\mathbf{B}_{\boldsymbol{\theta}}\boldsymbol{\beta}+\mathbf{Z}\boldsymbol{\alpha},\sigma^2[\mathbf{I}+\nu^2 \boldsymbol{P}_{\boldsymbol{\theta}}\mathbf{K}\boldsymbol{P}_{\boldsymbol{\theta}}])
		\end{array}
	\end{cases} \\
	\text{Priors}
	&\begin{cases}
		~~\begin{array}{r@{\hspace{3pt}}l}
			\boldsymbol{\beta}_j&\sim \gamma_j N(0,\tau^2_j \boldsymbol{S}_j^{-1})+(1-\gamma_j) \boldsymbol{\delta}_0, ~~~~\text{for $j=1,\cdots,p$  } \\
			\gamma_j &\sim \text{Bernoulli}(\pi) \\
			\pi & \sim \text{Beta}(a_{\pi},b_{\pi}) \\
			\tau_j^2 &\sim \text{Inv-Gamma}(a_{\tau},b_{\tau}) \\
			\boldsymbol{\theta}_j&|\text{max}(\gamma_j,\gamma_j^{\rho})=1\sim \text{vMF}(\kappa,\boldsymbol{\mu}) \\
			\rho_j&\sim \gamma_j^{\rho} \text{Gamma}(a^{\rho},b^{\rho})+(1-\gamma_j^{\rho})\boldsymbol{\delta}_0 \\
			\gamma_j^{\rho}&\sim \gamma_j\text{Bernoulli}(\pi^{\rho})+(1-\gamma_j)\boldsymbol{\delta}_0 \hfill ~~~~~~~~\text{ HIERARCHICAL}  \\
			\pi^{\rho} &\sim \text{Beta}(a_{\pi}^{\rho},b_{\pi}^{\rho}) \\
			\nu^{2} &\sim  \text{Inv-Gamma}(a_{*},b_{*}) \\
			\boldsymbol{\alpha} &\sim N(\mathbf{0},\mathbf{I}) \\
			\sigma^2 &\sim \text{Inv-Gamma}(a_{\sigma},b_{\sigma})
		\end{array}
	\end{cases}\\
	\text{Posterior}
	&\begin{cases}
		~~\begin{array}{l@{\hspace{3pt}}ll}
			|\boldsymbol{\Sigma}|^{-\frac{1}{2}}\exp\left(-\frac{1}{2}\left[ \mathbf{y}- (\mathbf{B}_{\boldsymbol{\theta}}\boldsymbol{\beta}+\mathbf{Z}\boldsymbol{\alpha}) \right]^T \boldsymbol{\Sigma}^{-1} \left[ \mathbf{y}- (\mathbf{B}_{\boldsymbol{\theta}}\boldsymbol{\beta}+\mathbf{Z}\boldsymbol{\alpha}) \right] \right),  &\quad \longrightarrow \quad\mathcal L(\cdot) \\
			\hfill \text{ where }\boldsymbol{\Sigma}=\sigma^2[\mathbf{I}_n+\nu^2 \boldsymbol{P}_{\boldsymbol{\theta}}\mathbf{K}\boldsymbol{P}_{\boldsymbol{\theta}}]  & \\
			~\times ~  \prod_{j=1}^{p} \left(\gamma_j (2\pi)^{-\frac{d_j}{2}} (\tau_j^2)^{-\frac{d_j}{2}} |\boldsymbol{S}_j|^{\frac{1}{2}} \exp(-\frac{\boldsymbol{\beta}_j^T\boldsymbol{S}_j\boldsymbol{\beta}_j}{2\tau_j^2})+[1-\gamma_j]\boldsymbol{\delta}_0\right)    &\quad \longrightarrow \quad f(\boldsymbol{\beta}) \\
			~\times ~ \pi^{[a_{\pi}+\sum_{j=1}^p \gamma_j-1]} (1-\pi)^{[b_{\pi}+p-\sum_{j=1}^p \gamma_j-1]} &\quad \longrightarrow \quad f(\gamma_j,\pi) \\
			~\times ~\frac{ b_{\tau}^{a_{\tau}} }{\Gamma(a_{\tau})} ( \tau_j^{2})^{-a_{\tau}-1} \exp\left(- \frac{b_{\tau}}{ \tau_j^{2}}\right)   &\quad \longrightarrow \quad f(\tau_j^{2})\\
			~\times ~\prod_{j=1}^p \exp \left[\{1-(1-\gamma_j)(1-\gamma_j^{\rho})\}\kappa \boldsymbol{\theta}_j^T \boldsymbol{\mu}\right] &\quad \longrightarrow \quad f(\boldsymbol{\theta}_j )\\
			~\times ~\prod_{j=1}^p \left(\gamma_j^{\rho}\frac{(b^{\rho} )^{a^{\rho} }}{\Gamma(a^{\rho} )}\rho_j^{a^{\rho}-1} \exp(-b^{\rho} \rho_j) + [1-\gamma_j^{\rho}]\boldsymbol{\delta}_0 \right) &\quad \longrightarrow \quad f(\rho_j )\\
			~\times ~ (\pi^{\rho})^{[a^{\rho}_{\pi}-1+\sum_{j=1}^p \gamma_j \gamma^{\rho}_j]} (1-\pi^{\rho})^{[b^{\rho}_{\pi}-1+\sum_{j=1}^p \gamma_j( 1-\gamma^{\rho}_j)]}  &\quad \longrightarrow \quad f(\gamma_j^{\rho},\pi^{\rho})  \\
			~\times ~ \frac{ b_{*}^{a_{*}} }{\Gamma(a_{*})} ( \nu^{2})^{-a_{*}-1} \exp\left(- \frac{b_{*}}{ \nu^{2}}\right)   &\quad \longrightarrow \quad f(\nu^{2})\\
			~\times ~ \exp\left( -\frac{1}{2}  \boldsymbol{\alpha}^T\boldsymbol{\alpha}  \right) &\quad \longrightarrow \quad f(\boldsymbol{\alpha}) \\
			~\times ~ \frac{b_{\sigma}^{a_{\sigma}}}{\Gamma(b_{\sigma})}(\sigma^2)^{-a_{\sigma}-1}\exp\left(-\frac{b_{\sigma}}{\sigma^2}\right) &\quad \longrightarrow \quad f(\sigma^2)
		\end{array}
	\end{cases}
\end{alignat*}

\clearpage
\subsection{MCMC Sampler}
\begin{enumerate}
	\item Between models move. For $j=1,\dots,p$, there are now 3 possible states for $(\gamma_j,\gamma^{\rho}_j)$: (0,0),  (1,0), (1,1). Given the current state  of $(\gamma_j,\gamma^{\rho}_j)$, randomly propose a move to one of the other two states. [Note that we randomly select from the two available moves at any state, so the  probability of selecting the $m^{th}$ move type (which includes the forward and reverse move) cancels out and we can then ignore it in the acceptance ratio. ]

	If the current state is $(\gamma_j,\gamma^{\rho}_j)=(0,0)$, this move involves increasing model dimensions because of the index $\boldsymbol{\theta}_j$; if the proposed state is $(\gamma_j,\gamma^{\rho}_j)=(0,0)$, then this move involves decreasing model dimensions, since the proposed model no longer includes $\boldsymbol{\theta}_j$. In both cases we need to use a formal MHG type move:
	\begin{itemize}
		\item Increasing dimension: If $\gamma_j=0$ and $\gamma_j^{prop}=1$, then propose $\boldsymbol{\beta}_j^{prop}$ from the slab prior, $N(\mathbf{0},\tau_j^2 \mathbf{S}_j^{-1})$. 	If $\gamma_j^{\rho, prop}={0}$, $\rho_j^{prop}=0$.  If $\gamma_j^{\rho}=0$ and $\gamma_j^{\rho,prop}=1$, then propose $\rho_j^{prop}$ from the slab prior, $\text{Gamma}(a^{\rho},b^{\rho})$. And $\boldsymbol{\theta}_j^{prop}=\mathbf{u}\sim vMF (\kappa, \boldsymbol{\mu})$. 
		Then draw $U\sim Unif(0,1)$ and accept $({\gamma_j},\boldsymbol{\beta}_j,{\gamma_j^{\rho}},{\rho}_j,\boldsymbol{\theta}_j)=({\gamma_j^{\rho,prop}},\boldsymbol{\beta}_j^{prop},{\gamma_j^{\rho,prop}},{\rho_j^{prop}},\boldsymbol{\theta}_j^{prop})$ if $$\hspace{-3.5cm}\log U < log\left( \frac{P({\gamma_j^{\rho,prop}},\boldsymbol{\beta}_j^{prop},{\gamma_j^{\rho,prop}},{\rho_j^{prop}},\boldsymbol{\theta}_j^{prop}|\mathbf{y},\dots)q({\gamma_j},\boldsymbol{\beta}_j,{\gamma_j^{\rho}},{\rho}_j|{\gamma_j^{\rho,prop}},\boldsymbol{\beta}_j^{prop},{\gamma_j^{\rho,prop}},{\rho_j^{prop}},\boldsymbol{\theta}_j^{prop})  }{
			P({\gamma_j},\boldsymbol{\beta}_j,{\gamma_j^{\rho}},{\rho}_j|\mathbf{y},\dots) q({\gamma_j^{prop}},\boldsymbol{\beta}_j^{prop},{\gamma_j^{\rho,prop}},{\rho_j^{prop}},\boldsymbol{\theta}_j^{prop}|{\gamma_j},\boldsymbol{\beta}_j,{\gamma_j^{\rho}},{\rho}_j)} \times \left|J\right| \right) $$
		where $|J|=1$ here. 
		
		Decreasing Dimension: Or if $(\gamma_j^{prop},\gamma_j^{\rho,prop})=(0,0)$, set $\boldsymbol{\beta}_j^{prop}=\mathbf{0}$, $\rho_j=0$, and  $\mathbf{u}'=\boldsymbol{\theta}_j$.		Then draw $U\sim Unif(0,1)$ and accept $({\gamma_j},\boldsymbol{\beta}_j,{\gamma_j^{\rho}},{\rho}_j)=({\gamma_j^{\rho,prop}},\boldsymbol{\beta}_j^{prop},{\gamma_j^{\rho,prop}},{\rho_j^{prop}})$ if $$\hspace{-3.5cm}\log U <log\left( \frac{P({\gamma_j^{\rho,prop}},\boldsymbol{\beta}_j^{prop},{\gamma_j^{\rho,prop}},{\rho_j^{prop}}|\mathbf{y},\dots)q({\gamma_j},\boldsymbol{\beta}_j,{\gamma_j^{\rho}},{\rho}_j,\boldsymbol{\theta}_j|{\gamma_j^{\rho,prop}},\boldsymbol{\beta}_j^{prop},{\gamma_j^{\rho,prop}},{\rho_j^{prop}})  }{
			P({\gamma_j},\boldsymbol{\beta}_j,{\gamma_j^{\rho}},{\rho}_j,\boldsymbol{\theta}_j|\mathbf{y},\dots) q({\gamma_j^{prop}},\boldsymbol{\beta}_j^{prop},{\gamma_j^{\rho,prop}},{\rho_j^{prop}}|{\gamma_j},\boldsymbol{\beta}_j,{\gamma_j^{\rho}},{\rho}_j,\boldsymbol{\theta}_j)}  \times \left|J\right| \right)$$
		where $|J|=1$ here. 
	\end{itemize}

	Alternatively if the current state \textit{and} the proposed state is one of  (1,0), or (1,1), then the dimension is maintained, and we can use a standard MH step:
	\begin{itemize}
		\item If $\gamma_j^{\rho, prop}=0$, set $\rho_j^{prop}=0$.  If $\gamma_j^{\rho}=0$ and $\gamma_j^{\rho,prop}=1$, then propose $\rho_j^{prop}$ from the slab prior, $\text{Gamma}(a^{\rho},b^{\rho})$.  Then draw $U\sim Unif(0,1)$ and accept $({\gamma_j},\boldsymbol{\beta}_j,{\gamma_j^{\rho}},{\rho}_j)=({\gamma_j^{\rho,prop}},\boldsymbol{\beta}_j^{prop},{\gamma_j^{\rho,prop}},{\rho_j^{prop}})$ if $$\hspace{-1.5cm}\log U <log\left( \frac{P({\gamma_j^{\rho,prop}},\boldsymbol{\beta}_j^{prop},{\gamma_j^{\rho,prop}},{\rho_j^{prop}}|{\boldsymbol{\gamma}_{-j}},\boldsymbol{\beta}_{-j},\boldsymbol{\gamma^{\rho}}_{-j},\boldsymbol{\rho}_{-j},\mathbf{y},\dots)q({\gamma_j},\boldsymbol{\beta}_j,{\gamma_j^{\rho}},{\rho}_j|{\gamma_j^{\rho,prop}},\boldsymbol{\beta}_j^{prop},{\gamma_j^{\rho,prop}},{\rho_j^{prop}})  }{
			P({\gamma_j},\boldsymbol{\beta}_j,{\gamma_j^{\rho}},{\rho}_j|{\boldsymbol{\gamma}_{-j}},\boldsymbol{\beta}_{-j},\boldsymbol{\gamma^{\rho}}_{-j},\boldsymbol{\rho}_{-j},\mathbf{y},\dots) q({\gamma_j^{\rho,prop}},\boldsymbol{\beta}_j^{prop},{\gamma_j^{\rho,prop}},{\rho_j^{prop}}|{\gamma_j},\boldsymbol{\beta}_j,{\gamma_j^{\rho}},{\rho}_j)} \right)$$
		(and $\boldsymbol{\beta}_j^{prop}=\boldsymbol{\beta}_j$ so you can ignore this above; just leaving there in case this changes.)
		\item Note: when we update $\boldsymbol{\beta}_j^{prop} \neq 0$, we cannot directly draw from the prior, since $\textbf{S}_j$ has a column/row of 0s (due to the unpenalized component). So we draw the final element of $\boldsymbol{\beta}_j^{prop}$ from $N(0,1)$.
	\end{itemize}

	\item Within model moves:
	\begin{enumerate}
		\item Index weights: 
		\begin{itemize}
			\item Refinement step to improve mixing: For all $j=1,\dots,p$ such that $max(\gamma_j,\gamma_j^{\rho})=1$, propose $\boldsymbol{\theta}_j\sim vMF(\kappa_{prop},\boldsymbol{\mu}_{prop})$, and we set $\boldsymbol{\mu}_{prop}$ to the current value, and set $\kappa_{prop}=1000$ (which is quite large, but worked well in Antoniadis et al. 2004)
		\end{itemize}
		\item Additive effects: 
		\begin{enumerate}
			\item Refinement step to improve mixing: draw $\boldsymbol{\beta}$ from full conditional
			\begin{align*}
			\boldsymbol{\beta}_{\gamma}&\sim {N}\left(\left[\mathbf{B}_{\gamma}^T\boldsymbol{\Sigma}^{-1}\mathbf{B}_{\gamma}+\mathbf{V}_{\gamma} \right]^{-1} \mathbf{B}_{\gamma}^T\boldsymbol{\Sigma}^{-1}\left[\mathbf{y}-\mathbf{Z}\boldsymbol{\alpha}\right], ~~ \left[\mathbf{B}_{\gamma}^T\boldsymbol{\Sigma}^{-1}\mathbf{B}_{\gamma}+\mathbf{V}_{\gamma} \right]^{-1}\right), \\
			\text{ where }\boldsymbol{\Sigma}&=\sigma^2[\mathbf{I}_n+\nu^2 \boldsymbol{P}\mathbf{K}\boldsymbol{P}],  \\
			\mathbf{V}_{\gamma} &= \sum_{k=1}^p \frac{\gamma_k}{\tau_k^2}\boldsymbol{S^*}_k, \\
			\text{and }~	\mathbf{B}_{\gamma} &\text{ contains columns corresponding to $x_k$ only if $\gamma_k=1$} 
			\end{align*}
			\item Draw $\tau_j^2$ from inverse gamma (full conditional) for $j=1,\dots,p$
			\begin{align*}
			\tau_j^2&\sim \text{Inv-Gamma}\left(a_{\tau}+\gamma_j \frac{d_j}{2} ,~~ b_{\tau} +\gamma_j \frac{1}{2} \boldsymbol{\beta}^T\boldsymbol{S^*}_j\boldsymbol{\beta}\right)
			\end{align*} (where $ \boldsymbol{\beta}^T\boldsymbol{S^*}_j\boldsymbol{\beta}= \boldsymbol{\beta}_j^T\boldsymbol{S}_j\boldsymbol{\beta}_j$)
			\item Draw  $\pi$ from Beta (full conditional):
			\begin{align*}
			\pi\sim \text{Beta}\left(a_{\pi}+\sum_{k=1}^p \gamma_k,~~  b_{\pi}+p-\sum_{k=1}^p \gamma_k\right)
			\end{align*}
		\end{enumerate}
		\item Non-additive effects: 
		\begin{enumerate}
			\item Refinement step to improve mixing:
			\begin{itemize}
				\item For all $j=1,\dots,p$ such that $\gamma_j^{\rho}=1$ propose $\rho_j^{prop}\sim \text{Gamma}(\frac{\rho_j^2}{s^2},rate=\frac{\rho_j}{s^2})$ where $s$ is some jump size (i.e. sample from a Gamma centered at the current value with some user-specified sd/jump size; default $s$=0.1). Then draw $U\sim Unif(0,1)$ and accept ${\rho}_j=\rho_j^{prop}$ if $\log U <log\left( \frac{P({\rho_j^{prop}}|\boldsymbol{\gamma^{\rho}},\boldsymbol{\rho}_{-j},\mathbf{y},\dots)q({\rho_j}|{\rho_j^{prop}})  }{
					P({\rho_j}|\boldsymbol{\gamma^{\rho}},\boldsymbol{\rho}_{-j},\mathbf{y},\dots) q({\rho_j^{prop}}|{\rho_j})} \right)$
			\end{itemize} 
			\item Draw $\pi^{\rho}$ from Beta (full conditional)
			\begin{align*}
			\pi^{\rho}\sim \text{Beta}\left(a_{\pi}^{\rho}+\sum_{j=1}^p  \gamma_j \gamma_j^{\rho},~~  b_{\pi}^{\rho}+\sum_{j=1}^p \gamma_j (1-\gamma_j^{\rho})\right)
			\end{align*}
			\item Draw $\nu^{2}$ from full conditional via Metropolis-Hastings step: propose $\nu^{2,prop}\sim \text{Gamma}\left(\frac{[\nu^{2}]^2}{s^2}, \frac{\nu^{2}}{s^2}\right)$, i.e. a Gamma centered at the current value with default sd/jump size $s$=0.1. Then draw $U\sim Unif(0,1)$ and accept $\nu^{2}=\nu^{2,prop}$ if $\log U <log\left( \frac{P(\nu^{2,prop}|\mathbf{y},\dots)q(\nu^{2}|\nu^{2,prop})  }{
				P(\nu^{2}|\mathbf{y},\dots) q(\nu^{2,prop}|\nu^{2})} \right)$
		\end{enumerate}
		\item Draw $\boldsymbol{\alpha}$ from Gaussian (full conditional)
		\begin{align*}
		\boldsymbol{\alpha}&\sim {N}\left(\left[\mathbf{Z}^T \boldsymbol{\Sigma}^{-1}\mathbf{Z}+\mathbf{I}_d \right]^{-1} \mathbf{Z}^T \boldsymbol{\Sigma}^{-1}\left[\mathbf{y}-\mathbf{B}\boldsymbol{\beta}\right], ~~ \left[\mathbf{Z}^T \boldsymbol{\Sigma}^{-1}\mathbf{Z}+\mathbf{I}_d \right]^{-1}\right) \\
		&\text{ where }\boldsymbol{\Sigma}=\sigma^2[\mathbf{I}_n+\nu^2 \boldsymbol{P}\mathbf{K}\boldsymbol{P}] 
		\end{align*}
		\item Draw $\sigma^2$ from inverse gamma (full conditional)
		\begin{align*}
		\sigma^2&\sim \text{Inv-Gamma}\left(a_{\sigma}+\frac{n}{2} ,~~ b_{\sigma} +\frac{1}{2}\left[ \mathbf{y}- (\mathbf{B}\boldsymbol{\beta}+\mathbf{Z}\boldsymbol{\alpha}) \right]^T [\mathbf{I}_n+\nu^2 \boldsymbol{P}\mathbf{K}\boldsymbol{P}] ^{-1} \left[ \mathbf{y}- (\mathbf{B}\boldsymbol{\beta}+\mathbf{Z}\boldsymbol{\alpha}) \right]\right)
		\end{align*}
	\end{enumerate}
\end{enumerate}

\clearpage
\subsection{Gaussian Predictive Process}
Inverting the $N \times N$ matrix $\boldsymbol{K}$ can be slow when $N$ is large. Instead we follow the Gaussian predictive process approach of \cite{bobb2018statistical} and \cite{savitsky2011variable}, by projecting onto a set of $N_1<N$ knots. Omitting the subscript $k$ here for simplicity, the projection approach involves first approximating $\boldsymbol{K}$  by $\boldsymbol{K}_{10}^T \boldsymbol{K}_{11}^{-1} \boldsymbol{K}_{10}$, where the joint kernel matrix for the user defined knots and the observed exposures is 
\begin{align*}
\left[\begin{array}{ll}
\boldsymbol{K}_{11} & \boldsymbol{K}_{10} \\
\boldsymbol{K}_{10}^T & \boldsymbol{K}
\end{array}\right].
\end{align*}

Then we apply the Woodbury identity to obtain the inverse matrix:
\begin{align*}
[\boldsymbol{I}+\tau^2 \mathbf{P}\boldsymbol{K}\mathbf{P}]^{-1}& \approx [\boldsymbol{I}+\tau^2 \mathbf{P}\boldsymbol{K}_{10}^T \boldsymbol{K}_{11}^{-1} \boldsymbol{K}_{10}\mathbf{P}]^{-1} \\
&=\boldsymbol{I}-\tau^2 \mathbf{P}\boldsymbol{K}_{10}^T [\boldsymbol{K}_{11}+\tau^2 \boldsymbol{K}_{10}\mathbf{P}\boldsymbol{K}_{10}^T]^{-1}\boldsymbol{K}_{10}\mathbf{P}
\end{align*}
and the matrix determinant lemma:
\begin{align*}
\log \text{det}[\boldsymbol{I}+\tau^2 \mathbf{P}\boldsymbol{K}\mathbf{P}]^{-1}&\approx \log \text{det}[\boldsymbol{I}+\tau^2 \mathbf{P}\boldsymbol{K}_{10}^T \boldsymbol{K}_{11}^{-1} \boldsymbol{K}_{10}\mathbf{P}]^{-1} \\
&=2tr \log \left(chol[\boldsymbol{K}_{11}] \right)-2tr \log \left(chol[\boldsymbol{K}_{11}+\tau^2 \boldsymbol{K}_{10}\mathbf{P}\boldsymbol{K}_{10}^T]\right)
\end{align*}

\clearpage
\subsection{Sign Flipping and Polar Transformation}
The main proposed approach we  only constrains $\boldsymbol{\theta}_j$ to have $L2$-norm one, following \cite{antoniadis2004bayesian}. This can sometimes lead to sign flipping without further constraint. One option is to constrain ${\theta}_{j1}>0$. Here we instead place uniform priors on the half unit-(hyper)-sphere. Sampling is somewhat more challenging in this case, so we transform the original weight vectors $\boldsymbol{\theta}_j$ to polar coordinates as in \cite{dhara2020new}: 
\begin{align*}
\theta_{j1}&= sin(\phi_{j1})\\
\theta_{j2}&= sin(\phi_{j2})cos(\phi_{j1})\\
& \vdots \\
\theta_{jL-1}&= sin(\phi_{jL-1})\prod_{l=1}^{L-2} cos(\phi_{jl})\\
\theta_{jL}&= \prod_{l=1}^{L-1} cos(\phi_{jl})
\end{align*}
where $\phi_{j1}\in[0,\pi/2]$ and $\phi_{jl}\in[-\pi/2,\pi/2]$ for $l=2,\dots,L-1$. This parameterization ensures $\theta_{j1}\geq 0$ in addition to the L2 constraint. Using this parameterization we can easily modify the sampler in two ways:
\begin{enumerate}
	\item Between-model moves: $\boldsymbol{\theta}_j^{prop}$ is again drawn from the prior when increasing dimension. The prior is now uniform on the unit half-sphere, so we draw from the uniform vMF and set  $\boldsymbol{\theta}_{j1}^{prop}= |\boldsymbol{\theta}_{j1}^{prop}|$.
	
	\item Within model-move
	\begin{enumerate}
		\item Index weights: Refinement step: For all $j=1,\dots,p$ such that $max(\gamma_j,\gamma^{\rho}_j)=1$, draw $\boldsymbol{\theta}^{prop}_j$ as follows: For $l=1,\dots,L_j-1$ draw $\phi^{prop}_{jl}$ using scaled/shifted Beta proposals: $ Beta(a_{\phi},b_{\phi})$ with the mode set to be the previous value; i.e. $b_{\phi}=((1-\left[\frac{\phi_{cj}+\frac{\pi}{2}}{\pi}\right])a_{\phi}+2 \left[\frac{\phi_{cj}+\frac{\pi}{2}}{\pi}\right]-1)/\left[\frac{\phi_{cj}+\frac{\pi}{2}}{\pi}\right]$ for $j>1$ $b_{\phi}=((1-\left[\frac{\phi_{cj}}{\pi/2}\right])a_{\phi}+2 \left[\frac{\phi_{cj}}{\pi/2}\right]-1)/\left[\frac{\phi_{cj}}{\pi/2}\right]$ for $j=1$. The acceptance ratio must then be corrected by adjusting for the change of variable  with factor $| \prod_{l=1}^{L-1}cos(\phi_{jl})^{L-j} |$.
	\end{enumerate} 
\end{enumerate}

\clearpage
\section{Additional Simulation Results}

\begin{table}[htbp!]
	\caption{Extended simulation results with $\sigma^2=1$: comparing proposed CKMR to BKMR, a non-linear interaction selection method (NLInter), a spike-and-slab GAM (ssGAM), and the MixSelect approach when data are generated from a componentwise model, with and without interaction. And comparing proposed CMIM to BMIM when data are generated from an indexwise model with and without interaction.} \label{supptab:simtab}
	\centering
	\begin{tabular}{llllrrrr}
		\toprule
	Scenario &	Indexwise & Interaction & Method & MSE & Bias & Width & Cvg  \\ 
		\midrule
		A & No & No 		&  ssGAM & 0.35 & -0.01 & 1.62 & 0.98 \\ 
		&&   & BKMR & 0.67 & -0.01 & 2.22 & 0.89 \\ 
		&&    &    NLInter & 0.44 & -0.00 & 1.68 & 0.94 \\ 
   &&  & MixSelect & 2.07 & -1.04 & 1.37 & 0.26 \\ 
		&&    &    NonAdaptive & 0.38 & -0.01 & 1.60 & 0.97 \\ 
		&&    &    CKMR & 0.38 & -0.01 & 1.60 & 0.96 \\ 
		\midrule
		B&No & Yes 		&  ssGAM & 0.69 & -0.01 & 1.91 & 0.84 \\ 
		&&    & BKMR & 0.72 & -0.00 & 2.36 & 0.88 \\ 
		&& &     NLInter & 0.69 & -0.00 & 2.09 & 0.87 \\ 
            &&  & MixSelect & 2.65 & -1.13 & 1.63 & 0.27 \\ 
		&& &     NonAdaptive & 0.53 & -0.01 & 1.97 & 0.94 \\ 
		&& &     CKMR & 0.49 & -0.01 & 1.87 & 0.95 \\ 
		
		\midrule
		C&Yes & No & BMIM & 0.68 & -0.01 & 2.46 & 0.92 \\ 
		&&    & NonAdaptive & 0.44 & -0.01 & 1.77 & 0.95 \\ 
		&&    & CMIM & 0.43 & -0.01 & 1.76 & 0.96 \\ 
		\midrule
		D&Yes& Yes & BMIM & 0.73 & -0.01 & 2.56 & 0.91 \\ 
		&&     & NonAdaptive & 0.65 & -0.01 & 2.18 & 0.91 \\ 
		&&     & CMIM & 0.56 & -0.01 & 2.08 & 0.94 \\ 
		\bottomrule
	\end{tabular}
\end{table}

\begin{table}[htbp!]
	\caption{Extended Simulations with $\sigma^2=2$: comparing proposed CKMR to BKMR, a non-linear interaction selection method (NLInter), a spike-and-slab GAM (ssGAM), and the MixSelect approach when data are generated from a componentwise model, with and without interaction. And comparing proposed CMIM to BMIM when data are generated from an indexwise model with and without interaction.} \label{supptab:simtab2}
	\centering
	\begin{tabular}{llllrrrr}
		\toprule
		Indexwise & Interaction & Method & MSE & Bias & Width & Cvg  \\ 
		\midrule
		A&No & No 		&  ssGAM & 0.48 & -0.01 & 2.15 & 0.97 \\ 
		&&   & BKMR & 0.81 & -0.01 & 2.72 & 0.89 \\ 
		&&    &    NLInter & 0.62 & -0.00 & 2.33 & 0.94 \\ 
   &&  & MixSelect & 2.23 & -1.21 & 1.48 & 0.26 \\ 
		&&    &    NonAdaptive & 0.51 & -0.02 & 2.18 & 0.97 \\ 
		&&    &    CKMR & 0.51 & -0.02 & 2.17 & 0.96 \\ 
		\midrule
		B&No& Yes 		&  ssGAM & 0.77 & -0.01 & 2.38 & 0.88 \\ 
		&&    & BKMR & 0.86 & -0.01 & 2.88 & 0.89 \\ 
		&& &     NLInter & 0.87 & -0.00 & 2.62 & 0.87 \\ 
  &&  & MixSelect & 2.78 & -1.24 & 1.70 & 0.26 \\ 
		&& &     NonAdaptive & 0.74 & -0.02 & 2.48 & 0.91 \\ 
		&& &     CKMR & 0.71 & -0.02 & 2.48 & 0.92 \\ 
		
		\midrule
		C&Yes & No & BMIM & 0.88 & -0.02 & 3.20 & 0.93 \\ 
		&&    & NonAdaptive & 0.59 & -0.02 & 2.40 & 0.96 \\ 
		&&    & CMIM & 0.59 & -0.02 & 2.40 & 0.96 \\ 
		\midrule
		D&Yes& Yes & BMIM & 0.91 & -0.01 & 3.26 & 0.92 \\ 
		&&     & NonAdaptive & 0.83 & -0.02 & 2.72 & 0.90 \\ 
		&&     & CMIM & 0.78 & -0.02 & 2.74 & 0.92 \\ 
		\bottomrule
	\end{tabular}
\end{table}

\begin{landscape}
	\begin{figure}[htbp!]
		\centering
		\includegraphics[width=0.49\linewidth]{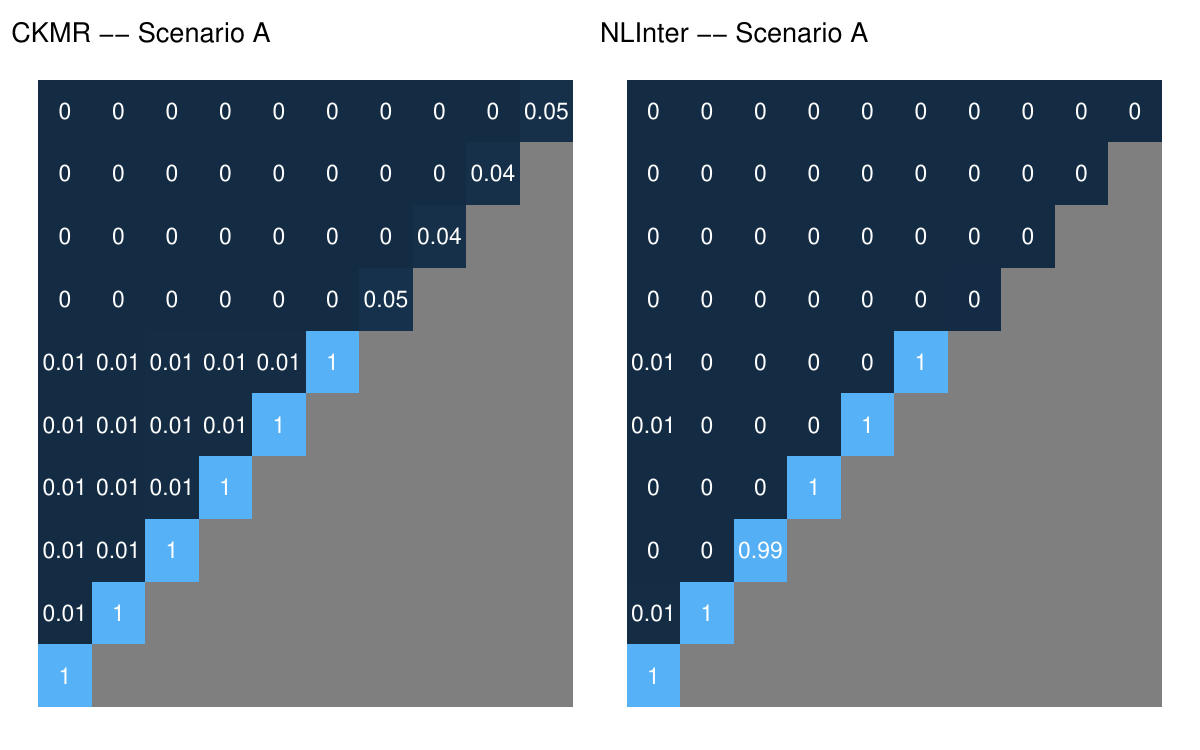}
		\includegraphics[width=0.49\linewidth]{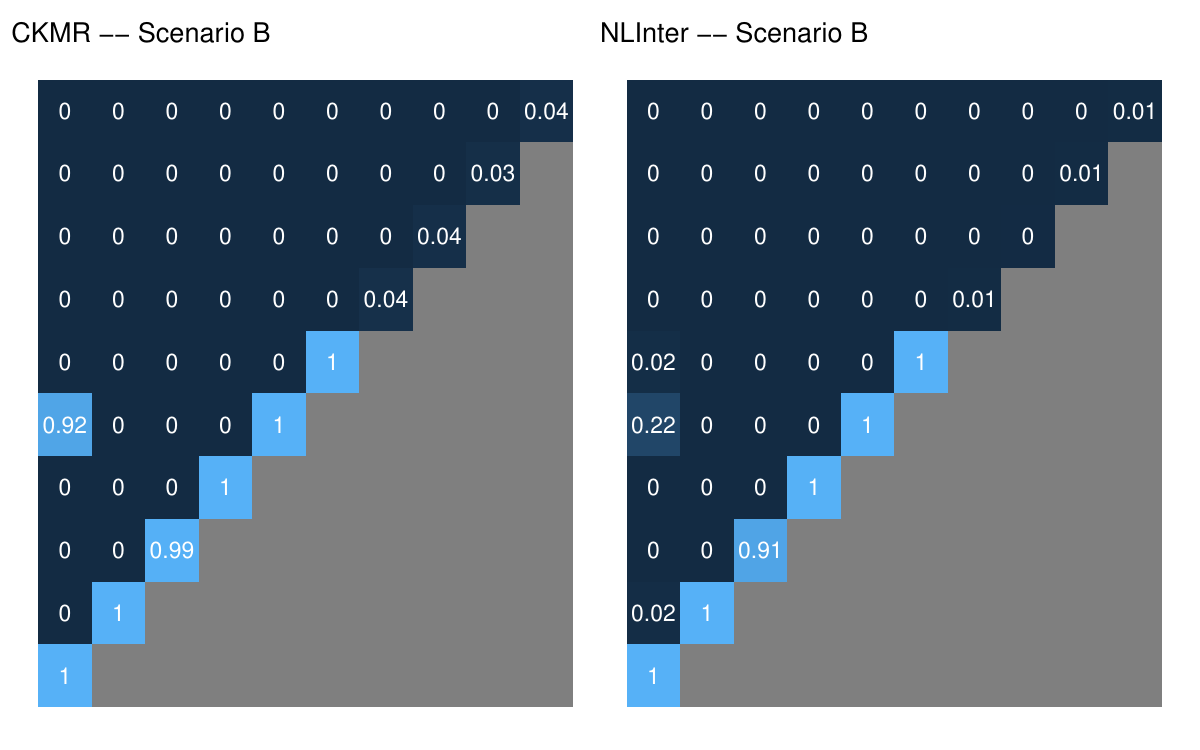} \\
		\includegraphics[width=0.49\linewidth]{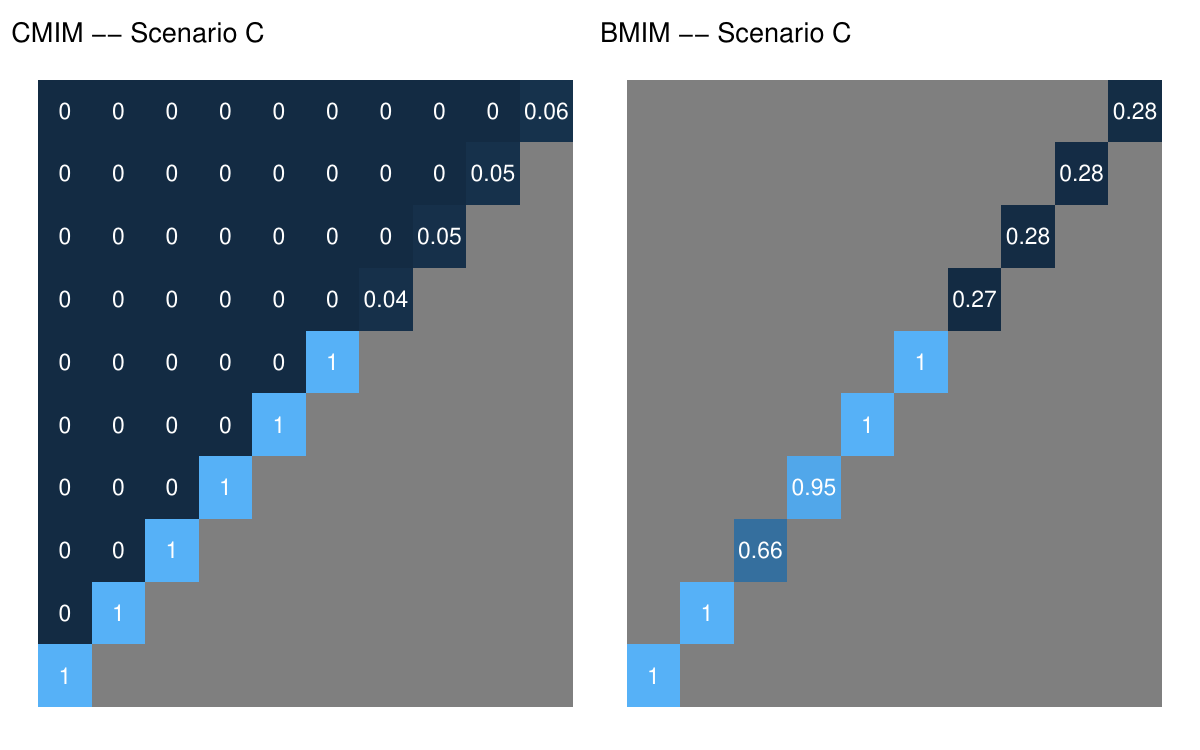}
		\includegraphics[width=0.49\linewidth]{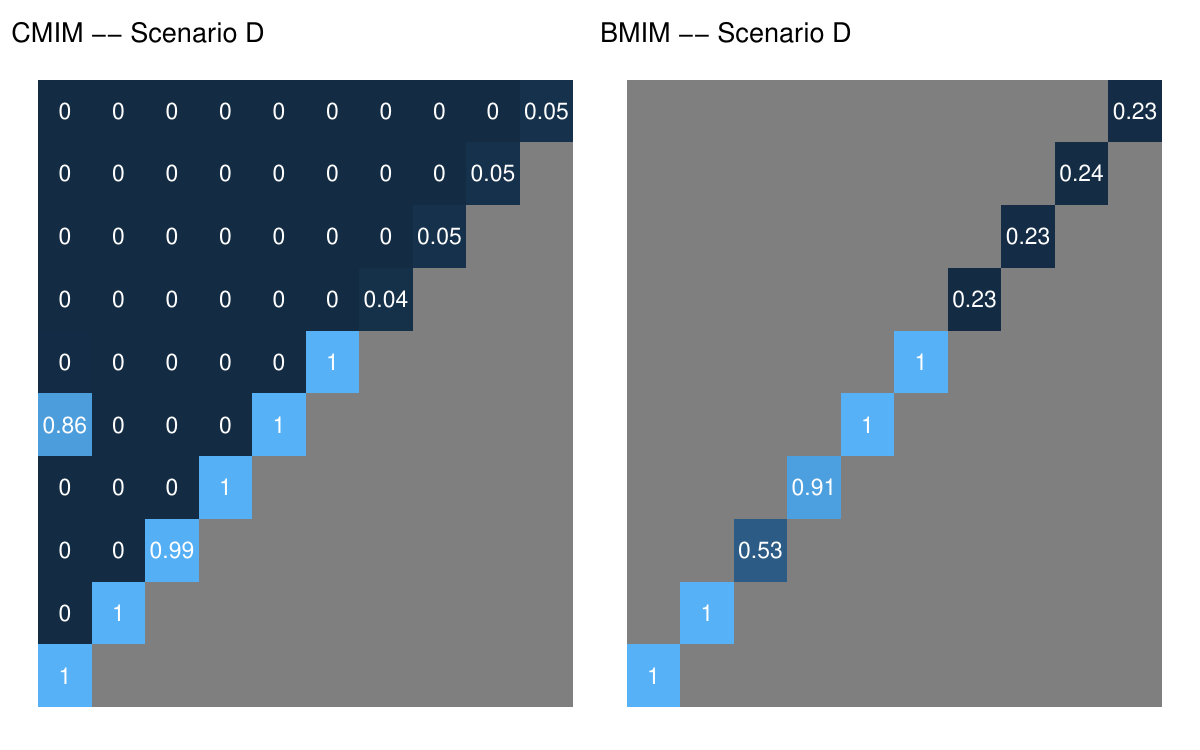}
		\caption{Heatplot of main+Interaction PIPs in componentwise (top row) and indexwise (bottom row) simulations with $\sigma^2=1$. First 6 components are non-null. In non-additive simulations (right two columns), components 1 \& 5 interact.}
		\label{fig:suppPIPs}
	\end{figure}
\end{landscape}

\begin{landscape}
	\begin{figure}[htbp!]
		\centering
		\includegraphics[width=0.49\linewidth]{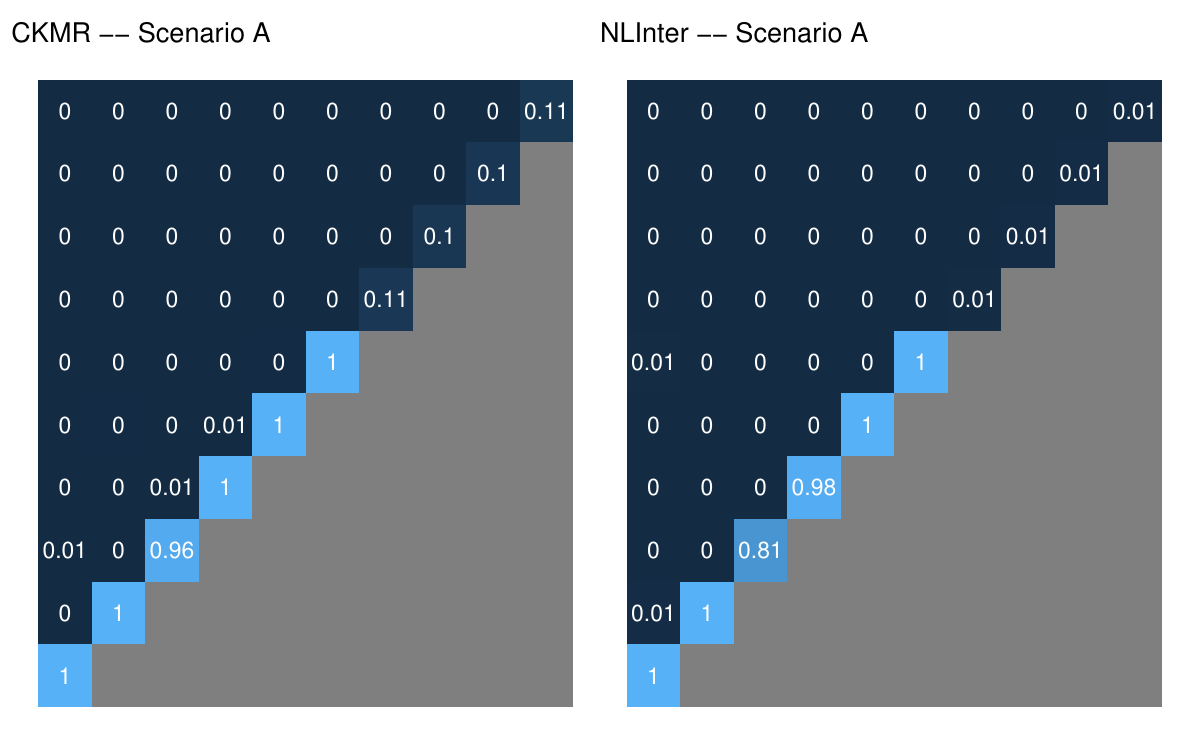}
		\includegraphics[width=0.49\linewidth]{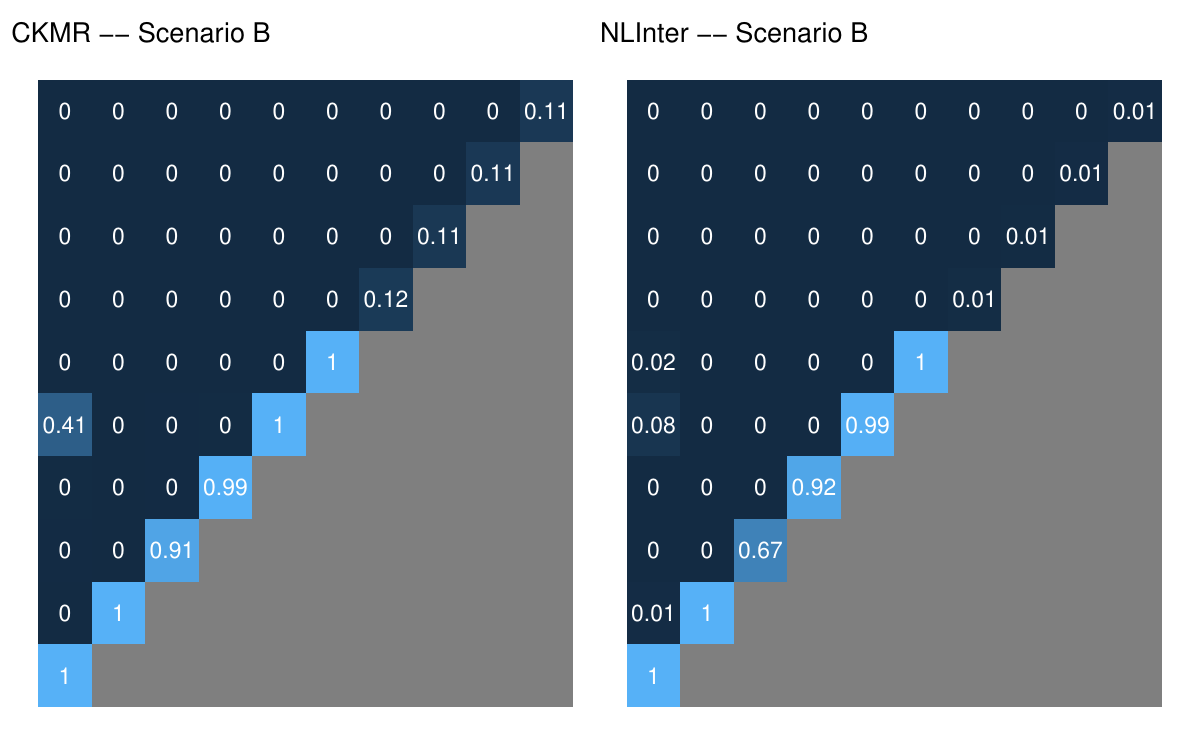} \\
		\includegraphics[width=0.49\linewidth]{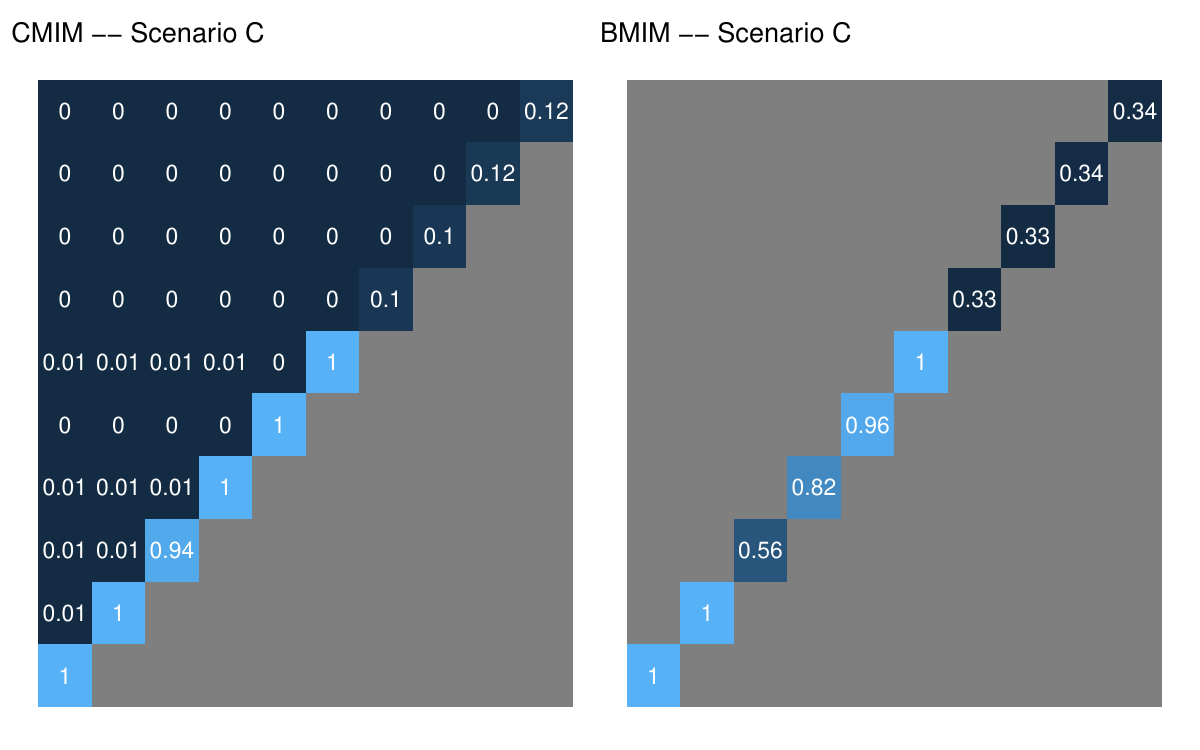}
		\includegraphics[width=0.49\linewidth]{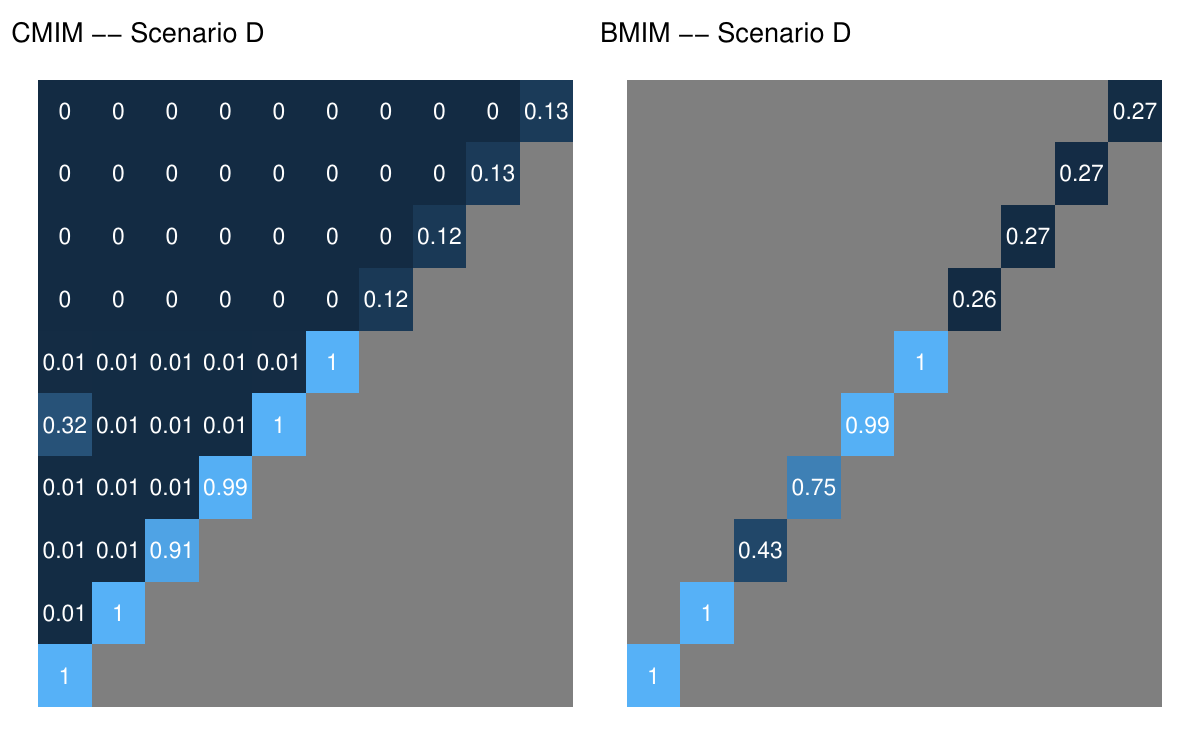}
		\caption{Heatplot of main+Interaction PIPs in componentwise (top row) and indexwise (bottom row) simulations with $\sigma^2=2$. First 6 components are non-null. In non-additive simulations (right two columns), components 1 \& 5 interact.}
		\label{fig:suppPIPs2}
	\end{figure}
\end{landscape}

\begin{figure}[htbp!]
	\centering
	\includegraphics[width=0.49\linewidth]{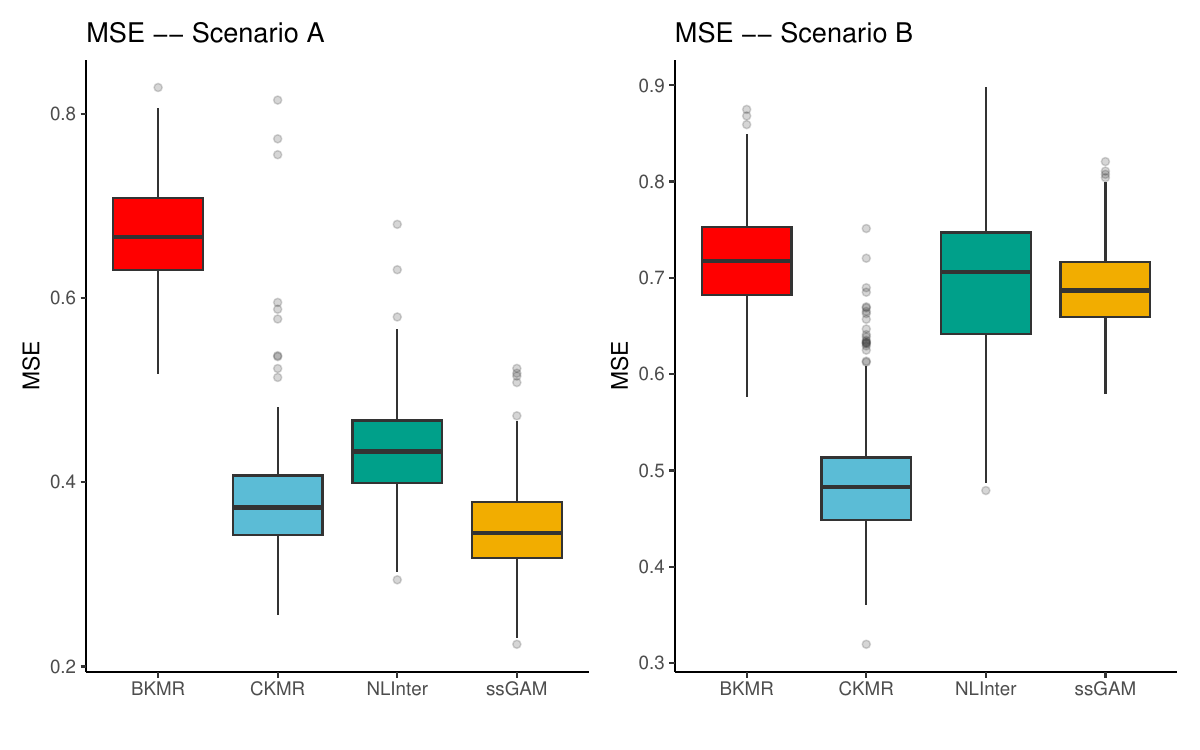} 
	\includegraphics[width=0.49\linewidth]{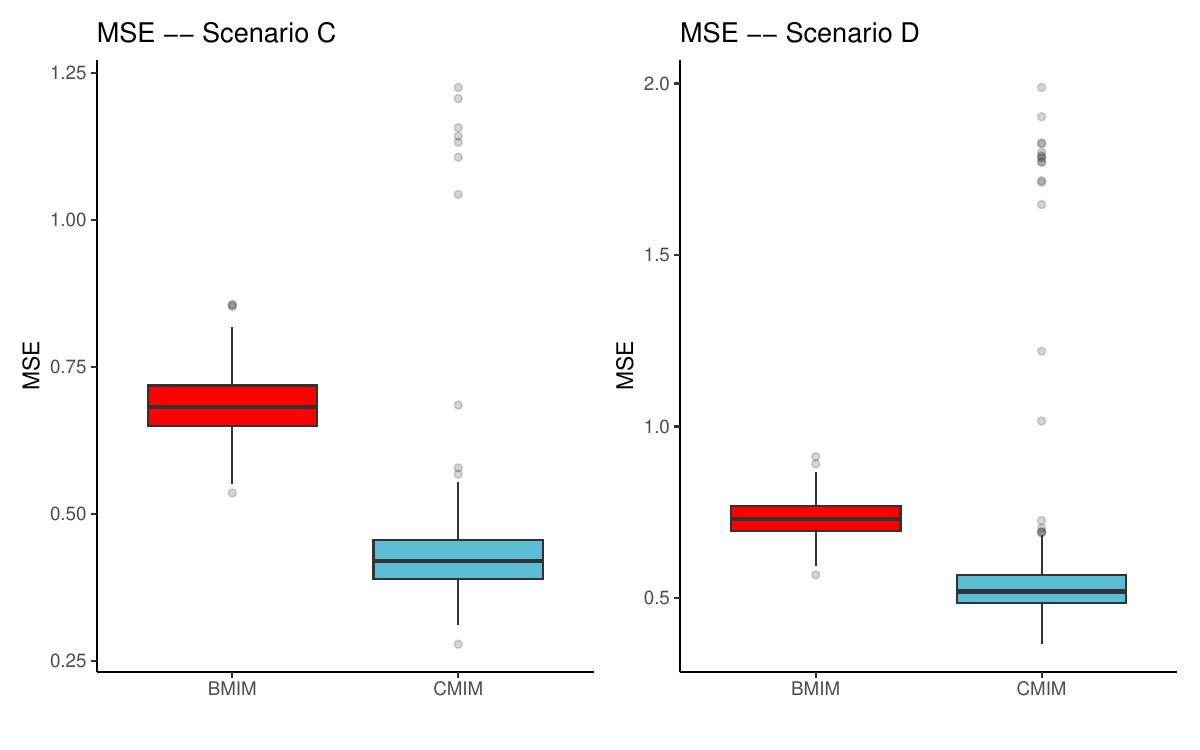} \\
		\includegraphics[width=0.49\linewidth]{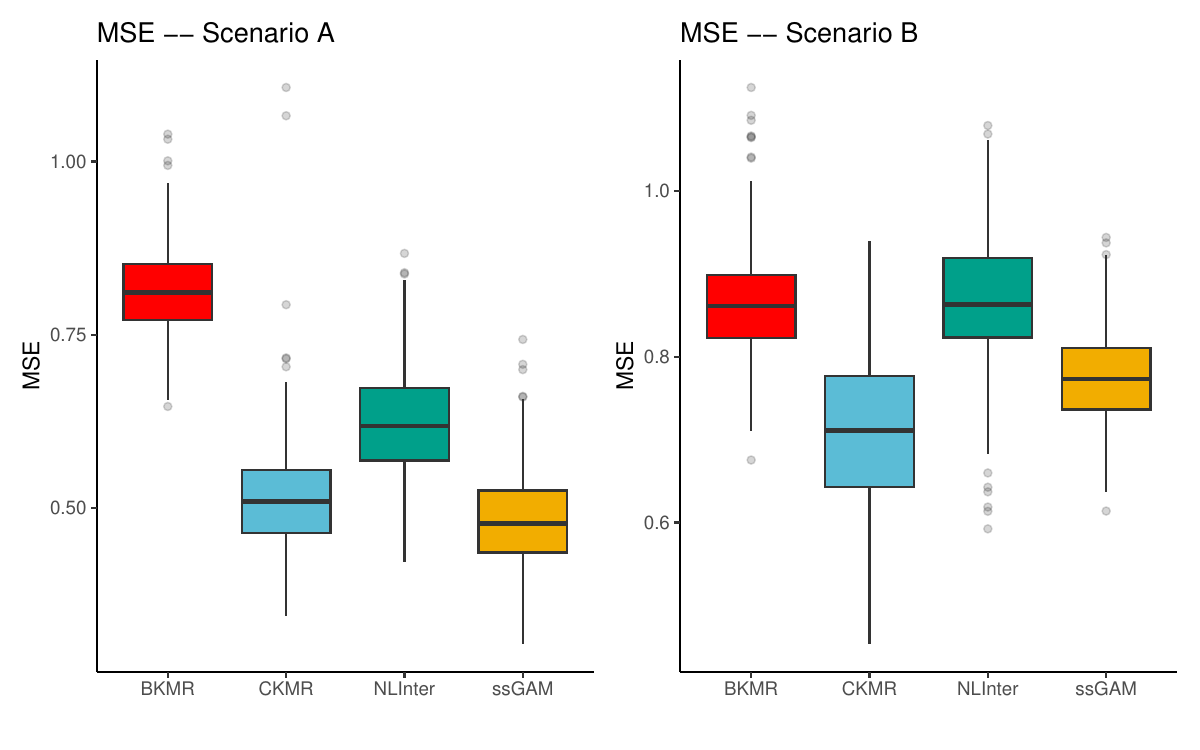} 
	\includegraphics[width=0.49\linewidth]{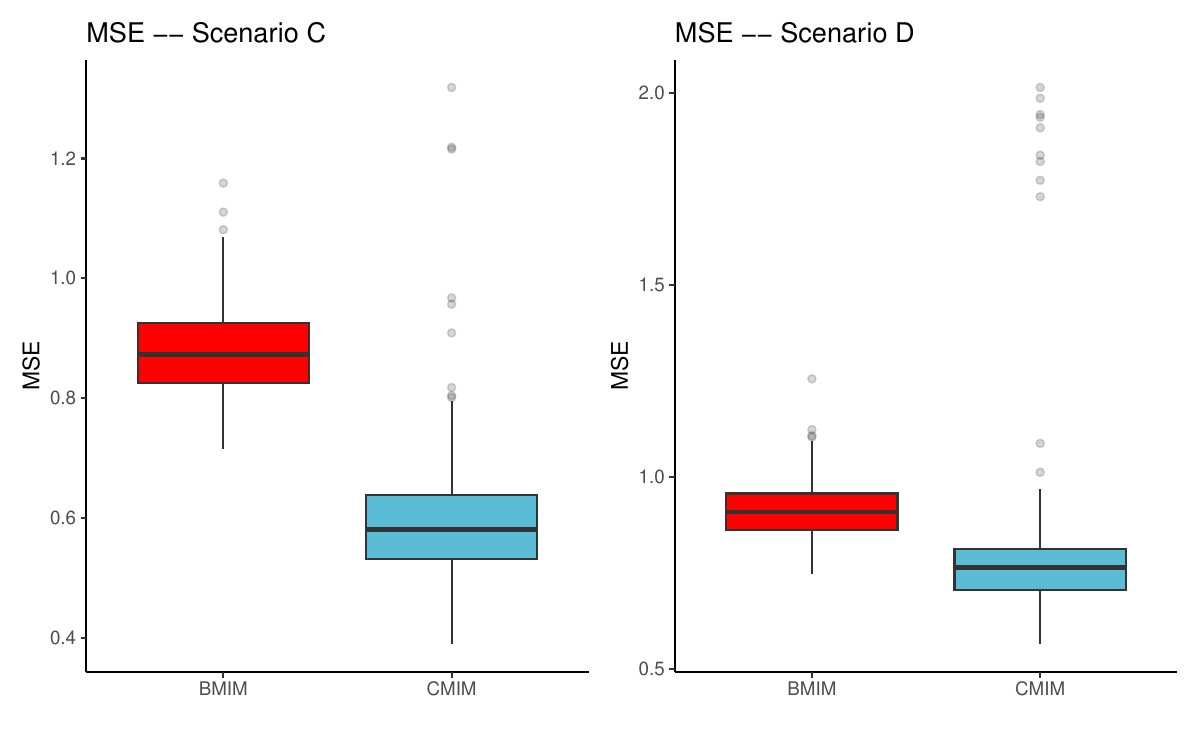}
	\caption{Distributions of of MSEs in componentwise (left 2 columns) and indexwise (right 2 columns) simulations with $\sigma^2=1$ (top row) and $\sigma^2=2$ (bottom row). In non-additive simulations (right two columns), components 1 \& 5 interact.}
	\label{fig:suppMSE}
\end{figure}
\clearpage
\subsection{Non-Adaptive Results}

	\begin{figure}[htbp!]
		\centering
		\includegraphics[width=0.79\linewidth]{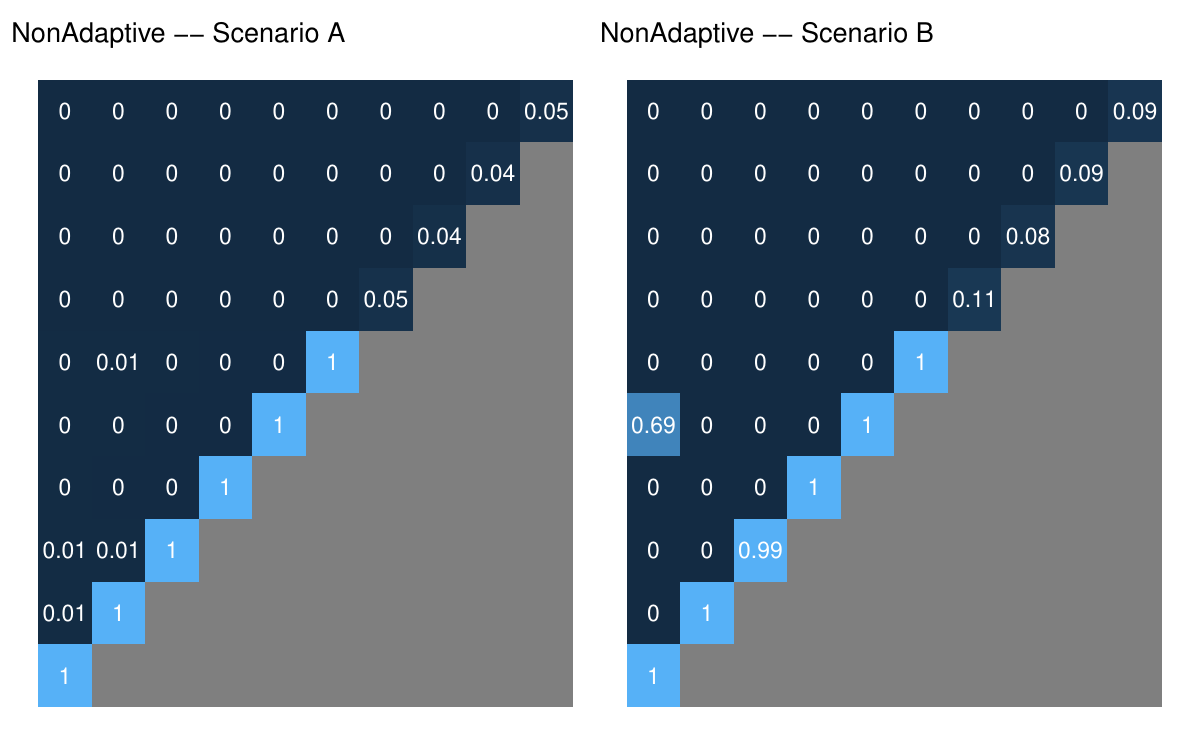}\\
		\includegraphics[width=0.79\linewidth]{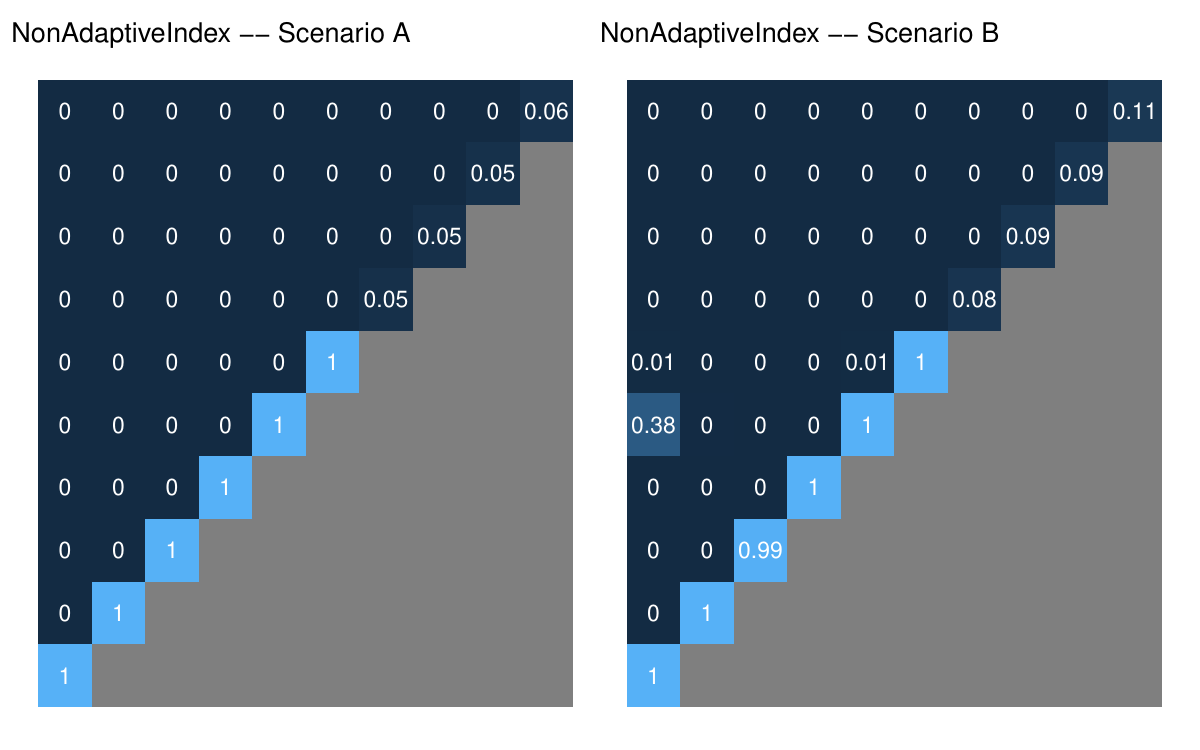} 
		\caption{Heatplot of main+Interaction PIPs in componentwise (top row) and indexwise (bottom row) simulations with $\sigma^2=1$. First 6 components are non-null. In non-additive simulations (right), components 1 \& 5 interact.}
		\label{fig:suppPIPsNonadaptive}
	\end{figure}

	\begin{figure}[htbp!]
	\centering
	\includegraphics[width=0.79\linewidth]{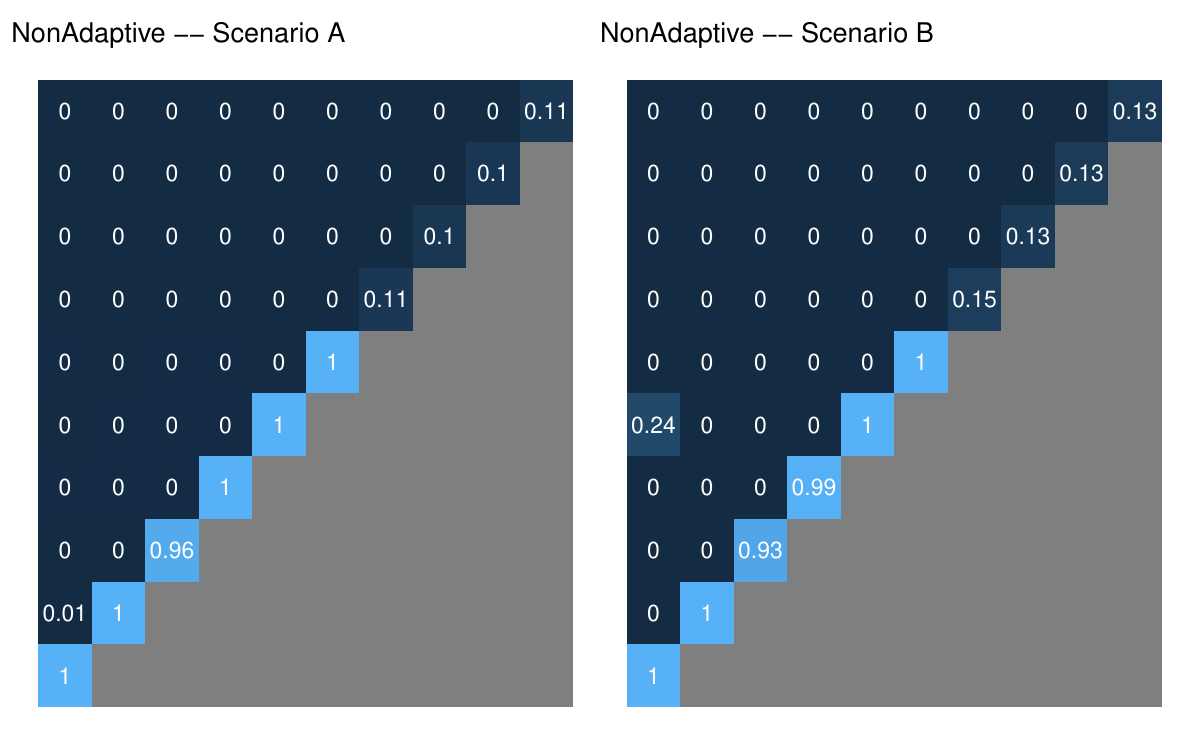}\\
	\includegraphics[width=0.79\linewidth]{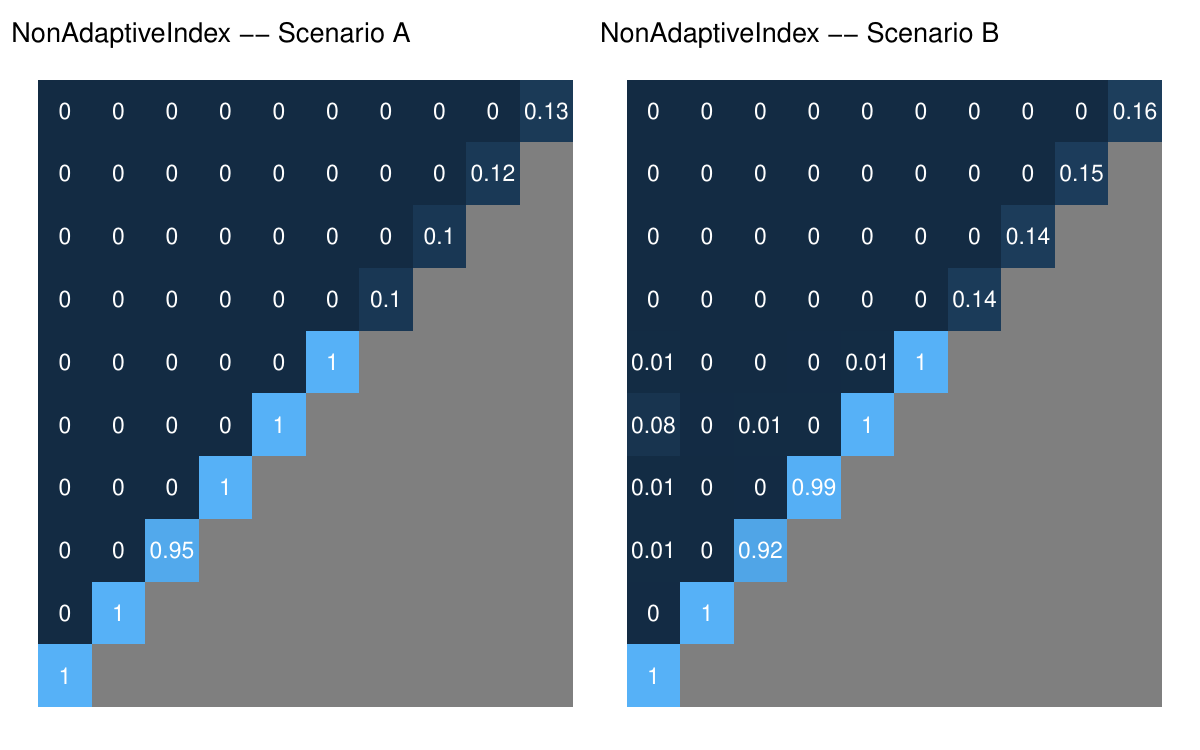} 
	\caption{Heatplot of main+Interaction PIPs in componentwise (top row) and indexwise (bottom row) simulations with $\sigma^2=2$. First 6 components are non-null. In non-additive simulations (right), components 1 \& 5 interact.}
	\label{fig:suppPIPsNonadaptive2}
\end{figure}

\clearpage
\section{Additional HELIX Results}


 \begin{table}[ht]
 \caption{HELIX Analysis: Indices, index size ($L_m$) and PIPs from CMIM.}
\centering
\begin{tabular}{lrr}
  \hline
  Index & $L_m$ & PIP \\ 
  \hline
 AirPollution &   4 & 22 \\ 
 Indoorair &   5 & 17 \\ 
 Lifestyle &   4 & 3 \\ 
 Metals &   9 & 100 \\ 
 Meteorological &   3 & 5 \\ 
 NaturalSpaces &   2 & 0 \\ 
 Organochlorines &   8 & 100 \\ 
 Organophosphates &   4 & 2 \\ 
 PBDE &   2 & 7 \\ 
 PFAS &   5 & 2 \\ 
 Phenols &   7 & 3 \\ 
 Phthalates &  10 & 3 \\ 
 Traffic &   2 & 0 \\ 
   \hline
\end{tabular}
\end{table}

\begin{table}[ht]
	\centering \caption{HELIX Analysis: 2.5th, 50th and 97.5th percentiles of the posteriors for index weights $\theta_{ml}$ among metals and organochlorines.}
	\begin{tabular}{rrrrr}
		\toprule & & \multicolumn{3}{c}{Posterior Percentiles for $\theta_{ml}$} \\
		\cmidrule{3-5}
		& Exposure  & \quad \quad ~~2.5\% & 50\% & 97.5\% \\ 
		\midrule
		&\emph{Metals}\\
		& As & -0.228 & 0.047 & 0.273 \\ 
		& Cd & -0.124 & 0.108 & 0.341 \\ 
		& Co & -0.088 & 0.159 & 0.404 \\ 
		& Cs & -0.733 & -0.522 & -0.240 \\ 
		& Cu & -0.704 & -0.491 & -0.219 \\ 
		& Hg & 0.018 & 0.301 & 0.539 \\ 
		& Mn & -0.104 & 0.150 & 0.414 \\ 
		& Mo & 0.182 & 0.390 & 0.608 \\ 
		& Pb & 0.000 & 0.262 & 0.491 \\ 
		%
		\midrule
		& \emph{Organochlorines}  \\
		& DDE & 0.135 & 0.274 & 0.418 \\ 
		& DDT & -0.124 & -0.019 & 0.090 \\ 
		& HCB & 0.777 & 0.886 & 0.941 \\ 
		& PCB118 & -0.210 & -0.076 & 0.056 \\ 
		& PCB138 & -0.140 & 0.023 & 0.310 \\ 
		& PCB153 & -0.105 & 0.098 & 0.308 \\ 
		& PCB170 & 0.176 & 0.295 & 0.454 \\ 
		& PCB180 & -0.125 & 0.015 & 0.156 \\ 
		\bottomrule
	\end{tabular}
\end{table}

\end{document}